\documentclass[12pt]{article}
\usepackage{amssymb,latexsym, amsmath,color,graphicx}
\newcommand{\rem}[1]{}
\newcommand{\comment}[1]{\vspace{5mm}\par
\framebox{\begin{minipage}[c]{.9 \textwidth} \color{blue}\rm #1
\end{minipage}}\vspace{5 mm}\par}
\newtheorem{theorem}{Theorem}[section]

\newtheorem{proposition}[theorem]{Proposition}

\newcommand{\beq}          {\begin{equation}}
\newcommand{\eeq}          {\end{equation}}
\newcommand{\beqa}          {\begin{eqnarray}}
\newcommand{\eeqa}          {\end{eqnarray}}
\newcommand{\sumin}          {\sum_{i=1}^N}
\newcommand{\sumijn}          {\sum_{i,j=1}^N}
\newcommand{\sumjn}          {\sum_{j=1}^N}
\newcommand{\half}        {\frac{1}{2}}

\begin{document}
\title{
Rotating Concentric Circular Peakons}
\author{
{\bf Darryl D. Holm}
\\Theoretical Division and Center for Nonlinear Studies
\\Los Alamos National Laboratory, MS B284
\\ Los Alamos, NM 87545
\\ email: dholm@lanl.gov
\and
Mathematics Department\\
Imperial College of Science, Technology and Medicine
\\ London SW7 2AZ, UK
\\ email: d.holm@imperial.ac.uk
\and
{\bf Vakhtang Putkaradze}
\\ Department of Mathematics and Statistics
\\ University of New Mexico  Albuquerque, NM 87131-1141
\\ email: putkarad@math.unm.edu
\and
and\\
{\bf Samuel N. Stechmann}
\\Theoretical Division and Center for Nonlinear Studies
\\Los Alamos National Laboratory, MS B284
\\ Los Alamos, NM 87545
\\ email: sam@t7.lanl.gov
\and
Courant Institute of Mathematical Sciences
\\New York University
\\ New York, NY 10012
\\ email: stechman@cims.nyu.edu
}

\maketitle

\begin{abstract}
We study invariant manifolds of measure-valued solutions of the partial
differential equation for geodesic flow of a pressureless fluid. These
solutions describe interaction dynamics on lower-dimensional support sets;
for example, curves, or filaments, of momentum in the plane.  The 2+1
solutions we study are planar generalizations of the 1+1 peakon solutions
of Camassa \& Holm [1993] for shallow water solitons. As an example, we
study the canonical Hamiltonian interaction dynamics of
$N$ rotating concentric circles of peakons, whose solution manifold is
$2N-$dimensional. Thus, the problem is reduced from infinite dimensions to
a finite-dimensional, canonical, invariant manifold. The existence of this
reduced solution manifold and many of its properties may be understood, by
noticing that it is also the momentum map for the action of diffeomorphisms
on the space of curves in the plane. We show both analytical and numerical
results.
\end{abstract} 
\tableofcontents
\section{Introduction and overview}

\subsection*{Geodesic flow in $n$ dimensions}
As first shown in Arnold [1966] \cite{Arnold[1966]}, Euler's equations of
ideal fluid dynamics represent geodesic motion on the volume-preserving
diffeomorphisms with respect to the  the $L^2$ norm of the velocity.
More generally, a time-dependent smooth map $g(t)$ is a
geodesic on the diffeomorphisms with respect to a kinetic energy norm
$KE=\frac{1}{2}\|\mathbf{u}\|^2$, provided its velocity, the
right-invariant tangent vector
$\mathbf{u}=\dot{g}g^{-1}(t)$, satisfies the {\bf vector Euler-Poincar\'e
equation},
\begin{equation}
\partial_t\mathbf{m}
=
-\,
\mathbf{u}\cdot\nabla\mathbf{m}
-
\nabla\mathbf{u}^T\cdot\mathbf{m}
-
\mathbf{m}{\rm\,div\,}\mathbf{u}
\equiv
-{\rm\,ad\,}^*_\mathbf{u}\mathbf{m}
\,.
\label{VectorEPeqn}
\end{equation}
Here ${\rm\,ad\,}^*$ is the adjoint with respect to $L^2$ pairing
$\langle\,\cdot\,,\,\cdot\,\rangle:
\mathfrak{g}^*\times\mathfrak{g}\to\mathbb{R}$ of the ad-action
(commutator) of vector fields $\mathbf{u},\mathbf{w}\in\mathfrak{g}$. That
is,
\begin{equation}
\langle{\rm\,ad\,}_\mathbf{u}^*\mathbf{m}\,,\,\mathbf{w}\,\rangle
=
-\,
\langle\,\mathbf{m}\,,{\rm\,ad\,}_\mathbf{u}\mathbf{w}\,\rangle
=
-\,
\langle\,\mathbf{m}\,,\,[\mathbf{u},\mathbf{w}]\,\rangle
\,.
\end{equation}
The momentum vector $\mathbf{m}\in\mathfrak{g}^*$ is defined as the
variational derivative of kinetic energy with respect to velocity,
\begin{equation}
\delta (KE)
=
\langle\,\mathbf{m}\,,\,\delta \mathbf{u}\,\rangle
\quad\Leftrightarrow\quad
\mathbf{m}
=\frac{\delta (KE)}{\delta \mathbf{u}}
\,.
\end{equation}
This defining relation for momentum closes the Euler-Poincar\'e equation
(\ref{VectorEPeqn}) for geodesic motion with respect to the kinetic energy
metric
$KE=\frac{1}{2}\|\mathbf{u}\|^2$. For more details, extensions and
applications of the Euler-Poincar\'e equation to both compressible and
incompressible fluid and plasma dynamics, see Holm, Marsden and Ratiu
[1998] \cite{HolmMarsdenRatiu[1998]}.

\subsection*{Geodesic flow with $H^1$ velocities in two dimensions}
In this paper, we consider the {\it solution behavior} of the
Euler-Poincar\'e equation (\ref{VectorEPeqn}) when the momentum vector is
related to the velocity by the two-dimensional Helmholtz operation,
\begin{equation}
\mathbf{m}=\mathbf{u}-\Delta\mathbf{u}
\,,\label{CHmom}
\end{equation}
where $\Delta$ denotes the Laplacian operator.
This Helmholtz relation arises when the kinetic energy is given by the
$H^1$ norm of the velocity,
\begin{equation}
KE
=
\frac{1}{2}\|\mathbf{u}\|_{H\,^1}^2
=
\frac{1}{2}
\int |\mathbf{u}|^2 + |\nabla\mathbf{u}|^2 \,dx\,dy
\,.
\label{H1-norm}
\end{equation}
The $H^1$ kinetic energy norm (\ref{H1-norm}) is an approximation of the
Lagrangian in Hamilton's principle for columnar motion of shallow water
over a flat bottom, when potential energy is negligible (the zero linear
dispersion limit) and the kinetic energy of vertical motion is approximated
by the second term in the integral, \cite{CamassaHolmLevermore[1996]}. In
this approximation, the physical meaning of the quantity
$\mathbf{m}=\delta(KE)/\delta\mathbf{u}$ in the Helmholtz relation
(\ref{CHmom}) is the momentum of the shallow water flow, while
$\mathbf{u}$ is its velocity in two dimensions. See Kruse et al. [2001]
\cite{Kruse-etal[2001]} for
details of the derivation of the geodesic equation for approximating 2D
shallow water dynamics in this limit.

\subsubsection*{Problem statement: Geodesic flow on $H^1$ in cylindrical
symmetry}  The present work studies azimuthally symmetric solutions of
the Euler-Poincar\'e equation (\ref{VectorEPeqn}) in polar coordinates
$(r,\phi)$,
\beq
\mathbf{m}=m_r(r,t)\,\mathbf{\hat{r}}
+m_\phi(r,t)\,\boldsymbol{\hat{\phi}}
\quad\hbox{and}\quad
\mathbf{u}=u_r(r,t)\,\mathbf{\hat{r}}
+u_\phi(r,t)\,\boldsymbol{\hat{\phi}}
\,.
\label{polar-soln}
\eeq
In the standard basis for cylindrical coordinates, momentum one-forms and
velocity vector fields are expressed as,
\beq
\mathbf{m}\cdot\mbox{d}\mathbf{x}=m_r\,\mbox{d}r + rm_\phi\,\mbox{d}\phi
\quad\hbox{and}\quad
\mathbf{u}\cdot\nabla=u_r\partial_r + \frac{u_\phi}{r}\partial_\phi
\,.
\label{polar-vect+oneform}
\eeq
Solutions (\ref{polar-soln}) satisfy coupled partial differential equations
whose radial and azimuthal components are, respectively,
\beqa
\frac{\partial m_r}{\partial t}
&=&
-\, \frac{1}{r}\partial_r\big(rm_ru_r\big)
- m_r\partial_ru_r
- (r m_\phi)\partial_r \Big(\frac{u_\phi}{r}\Big)
\,,\label{mr-intro}
\\
\frac{\partial (rm_\phi)}{\partial t}
&=&
-\, \frac{1}{r}\partial_r\big(r^2m_\phi u_r\big)
\,.\label{mphi-intro}
\eeqa
In these coupled equations, nonzero rotation $u_\phi$ generates radial
velocity $u_r$, which influences the azimuthal motion.
Without rotation, $u_\phi=0$, and the solution becomes purely radial.
The system of equations (\ref{mr-intro}) and (\ref{mphi-intro}) for
geodesic motion conserves the $H^1$ kinetic energy norm,
\beq
K\!E([u_r],[u_\phi])
=
\frac{1}{2}\int \Big[
u_\phi^2 + u_{\phi,r}^2 + \frac{u_\phi^2}{r^2}
+
u_r^2 + u_{r,r}^2 + \frac{u_r^2}{r^2}
\Big] r \mbox{d} r
=
\frac{1}{2}\|\mathbf{u}\|^2_{H\,^1}
\,.
\label{erg-intro}
\eeq
The corresponding momenta satisfy the dual relations,
\beq
rm_\phi
=
\frac{\delta K\!E}{\delta (u_\phi/r)}
\quad\hbox{and}\quad
m_r
=
\frac{\delta K\!E}{\delta u_r}
\label{mom-def}
\,.
\eeq
The Helmholtz relation (\ref{CHmom}), between $m_\phi$ and
$u_\phi$, for example, becomes
\beq
m_\phi
=
u_\phi-\frac{\partial^2 u_\phi}{\partial r^2}
- \frac{1}{r}\frac{\partial u_\phi}{\partial r}
+ \frac{1}{r^2}u_\phi
=
\Big(1 - \frac{1}{r} \partial_r r \partial_r + \frac{1}{r^2}\Big)u_\phi
\label{helmoltzr}
\,.
\eeq
This relation between velocity and momentum defines the symmetric invertible
Helmholtz operator in cylindrical geometry with finite boundary conditions
at $r=0$ and $\to\infty$. The velocity $u(r,t)$ is obtained from the
momentum $m(r,t)$ by the convolution
$u(r)=G\ast m=\int_0^\infty G(r,\xi) m(\xi)\xi\mbox{d} \xi$ (an extra factor
$\xi$ arises in  cylindrical geometry) with the Green's function
$G(r,\xi)$.  The Green's function for the radial Helmholtz operator in
(\ref{helmoltzr}) is given by,
\beq
G(r,\xi)=\left\{ \begin{array}{cc}
I_1(\xi) K_1(r) \quad\hbox{for}\quad & \xi< r\,, \\
I_1(r) K_1( \xi) \quad\hbox{for}\quad & r< \xi\,,
\end{array}
\right.
\label{greensfcn}
\eeq
where $I_1$ and $K_1$ are modified Bessel's functions.
This Green's function will play a significant role in what follows.

Equations (\ref{mr-intro}) and (\ref{mphi-intro}) may be written
equivalently as a Hamiltonian system, by Legendre transforming the kinetic
energy $K\!E([u_r],[u_\phi]) $ to the Hamiltonian
$h([m_r],[rm_\phi])$, after which the equations take the form,
\begin{equation}\label{Ham-matrix-diff}
\frac{\partial}{\partial t}
     \begin{bmatrix}
     m_r \\ rm_\phi
     \end{bmatrix}
=
     \begin{bmatrix}
     \{m_r,h\} \\ \{rm_\phi,h\}
     \end{bmatrix}
= -\, \mathcal{D}
    \begin{bmatrix}
    {\delta h/\delta m_r} \\
    {\delta h/\delta (rm_\phi)}
    \end{bmatrix}
= -\, \mathcal{D}
     \begin{bmatrix}
     u_r \\ u_\phi / r
     \end{bmatrix}
\,,
\end{equation}
where ${\delta h/\delta m_r}=u_r$, ${\delta h/\delta (rm_\phi)}=u_\phi / r$
and the Hamiltonian operator $\mathcal{D}$,
\begin{equation}
\mathcal{D}=
    \begin{bmatrix}
    r^{-1}\partial_r r m_r + m_r\partial_r &
    rm_\phi\partial_r &
    \\
    r^{-1}\partial_r r^2 m_\phi & 0
    \end{bmatrix}
\,,
\label{Ham-op}
\end{equation}
is the matrix $\mathcal{D}=\{(m_r,rm_\phi)\,,\,(m_r,rm_\phi)\}$, which
defines the  Lie-Poisson bracket for geodesic motion in cylindrical
geometry. Note that $\mathcal{D}$ is skew-symmetric with respect to the
$L^2$ pairing with cylindrical radial measure.

\rem{
Before proceeding further with our study of the dynamics of
peakon solutions of (\ref{VectorEPeqn}) in the cylindrical case, we shall
first review the relevant solution properties of the one-dimensional
Euler-Poincar\'e equation (\ref{VectorEPeqn}) on the real line.

\subsection*{One-dimensional geodesic flows on the real line}
The one-dimensional version of the Euler-Poincar\'e equation
(\ref{VectorEPeqn}) on the real line is
\begin{equation}
m_t
+
um_x
+
2mu_x
=0
\,,
\quad\hbox{with}\quad
m=u-u_{xx}
\,,
\label{1dCH}
\end{equation}
with subscript notation for partial derivatives in time $t$ and position
$x$, and boundary conditions $u=0$ and $u_x=0$ as $|x|\to\infty$. This
equation was introduced in Camassa and Holm [1993] as the zero linear
dispersion limit for a model of shallow water motion obtained by using
Hamiltonian asymptotics \cite{CamassaHolm[1993]}. As shown in
Dullin, Gottwald and Holm [2001] \cite{DullinGottwaldHolm[2001]}, the
one-dimensional CH equation (\ref{1dCH}) also emerges in the limit of zero
linear dispersion at {\it quadratic order} from the same asymptotic
expansion of Euler's equations for which the famous Korteweg-de Vries (KdV)
soliton equation for shallow water waves arises at  {\it linear order}.

\subsection*{Shallow water solitons (peakons) on the real line}
Amongst other results, Camassa and Holm [1993] \cite{CamassaHolm[1993]}
showed the following properties of the CH equation that are relevant in the
present work:
\begin{enumerate}
\item\label{CH-int}
Like  KdV, the 1D CH equation (\ref{1dCH}) is completely
integrable as a  bi-Hamiltonian system.
\item\label{CH-evp}
\rem{
Via the Gelfand-Dorfman squared-eigenfunction construction, its
bi-Hamiltonian property implies that CH (\ref{1dCH}) may be re-formulated
as a compatibility condition for two linear equations. One of these is a
second-order isospectral eigenvalue problem, related to the classical
equation for the eigenmodes of a linear string. Hence, the
}

The CH equation (\ref{1dCH}) may be solved by the inverse scattering
transform method for a second-order isospectral eigenvalue problem,
related to the classical equation for the eigenmodes of a linear string.
Thus, spectral analysis of the linear string provides the eigenvalues of
the CH isospectral problem and, hence, the asymptotic speeds of its
solitons.
\item\label{CH-zdl-tw-soliton}
In the zero linear dispersion limit that we also consider here, the
isospectrum of the CH solitons for a spatially confined initial momentum
distribution is {\it purely discrete}. The single soliton solution of
the 1D CH equation (\ref{1dCH}) in this limit is the traveling wave,
$u(x,t)=ce^{-|x-ct|}$. Thus, the CH soliton travels with a speed equal to
its maximum height, at which its velocity profile has a sharp peak in the
zero linear dispersion limit. The peaked CH soliton is called a
{\bf peakon}.
\item\label{CH-Npkn}
The CH equation (\ref{1dCH}) possesses a finite-dimensional
invariant manifold given by the superposed sum of $N-$peakon solutions,
\beq
u(x,t)=\sum_{i=1}^N p_i(t) e^{-|x-q_i(t)|}
\,,\hbox{ for which }
m(x,t)=\sum_{i=1}^N 2p_i(t) \delta(x-q_i(t))
\,,
\label{Npkn-soln}
\eeq
where $G(x)=e^{-|x|} $ is the Green's function for the one-dimensional
Helmholtz operator, as $(1-\partial^2)e^{-|x|}=2\delta(x)$.
\item\label{CH-pq-dyn}
The peakon solutions of the 1D CH equation (\ref{1dCH}) superpose as the
linear sum in the $N-$peakon solution (\ref{Npkn-soln}), {\it provided}
the parameters
$p_i(t)$ and $q_i(t)$ satisfy
Hamilton's canonical equations
$\dot{p}_i(t)=-\partial H/\partial q_i$ and
$\dot{q}_i(t)=\partial H/\partial p_i$, with Hamiltonian,
\beq
H(p,q)=\half \sum_{i,j=1}^N p_i p_j e^{-|q_i-q_j|}
\label{1dham}
\eeq
Thus, peakon dynamics for CH is geodesic motion in $2N-$dimensional phase
space of the parameters $(p_i,q_i)$ with co-metric given by
$e^{-|q_i-q_j|}$.
\item\label{CH-decomp}
Any spatially confined initial momentum distribution splits up into these
elementary peakon solutions, whose interactions occur by elastic
collisions whose two-body phase shifts may be computed analytically from
this geodesic motion. Thus, the peakon solutions dominate the initial value
problem for the dispersionless CH equation on the real line.
\end{enumerate}

\subsection*{Extensions of CH on the real line}
\subsubsection*{The $b-$equation and peakons}
The one-dimensional version of (\ref{VectorEPeqn}) for $m(x,t)$ has now been
exhaustively studied \cite{1Drefs} and several extensions of it are also
available in the literature. For example, the equation
\begin{equation}
m_t
+
um_x
+
b\,mu_x
=0
\,,
\quad\hbox{with}\quad
m=u-u_{xx}
\,,
\label{1d-beqn}
\end{equation}
with a real parameter $b$ also has $N-$peakon solutions (\ref{Npkn-soln}),
which were shown to be stable numerically for $b>1$ in Holm and Staley
[2003] \cite{HolmStaley[2003a]}. The case $b=3$ was studied in Degasperis,
Holm and Hone [2002] \cite{DegasperisHolmHone[2002]} and was shown to be
also completely integrable, by a cubic isospectral problem, although
the $(p,q)-$dynamics for its $N-$peakon solutions (\ref{Npkn-soln}) is not
canonical. For more details about the solution dependence on the parameter
$b$ in equation (\ref{1d-beqn}), see Holm and Staley [2003a,2003b],
\cite{HolmStaley[2003a]}, \cite{HolmStaley[2003b]}. As shown in
Dullin, Gottwald and Holm [2001] \cite{DullinGottwaldHolm[2001]}, the
$b-$equation (\ref{1d-beqn}) is asymptotically equivalent to the
dispersionless CH equation (\ref{1dCH}) under a certain group of normal
form transformations, provided $b\ne-1$.

\subsubsection*{The $b-$equation and pulsons on the real line}
In another example of special relevance here, Fringer and Holm [2001]
\cite{FringerHolm[2001]} showed that the pulson equation for $b=2$  on the
real line,
\begin{equation}
m_t
+
um_x
+
2mu_x
=0
\,,
\quad\hbox{with}\quad
u(x)=\int_{-\infty}^\infty G(|x-y|)\,m(y)\,dy
\,,
\label{1dFH}
\end{equation}
preserves CH properties 4-6 above, but {\it without} requiring
integrability. In particular, the $N-$pulson solutions of equation
(\ref{1dFH}) are:
\beq
u(x,t)=\sum_{i=1}^N p_i(t) G(|x-q_i(t)|)
\,,\hbox{ for which }
m(x,t)=\sum_{i=1}^N p_i(t) \delta(x-q_i(t))
\,.
\label{Nplsn-soln}
\eeq
Just as for the peaked solitons of the integrable CH equation, these
$N-$pulson solutions dominate the dynamics of the initial value problem for
equation (\ref{1dFH}) on the real line. Of special relevance to the
present investigation, the parameters
$p_i(t)$ and $q_i(t)$ in the $N-$pulson  solutions (\ref{Nplsn-soln})
are found to satisfy Hamilton's canonical equations
$\dot{p}_i(t)=-\partial H/\partial q_i$ and
$\dot{q}_i(t)=\partial H/\partial p_i$, with Hamiltonian,
\beq
H(p,q)=\half \sum_{i,j=1}^N p_i p_j G(|q_i-q_j|)
\,.
\label{1dham}
\eeq
Thus, $N-$pulson dynamics for the pulson equation (\ref{1dFH}) is
again geodesic motion in $2N-$dimensional phase space of the parameters
$(p_i,q_i)$. The co-metric for this motion is given by the Green's
function $G(|q_i-q_j|)$, which reduces to the peakon case when
$G(x)=e^{-|x|}$. Any spatially confined initial momentum distribution
splits up into these elementary pulson solutions, whose interactions occur
by elastic collisions. The  two-body phase shifts for these collisions may
be computed analytically for this geodesic motion, for any symmetric
Green's function.

The corresponding two-pulson interactions for the pulson $b-$equation,
\begin{equation}
m_t
+
um_x
+
b\,mu_x
=0
\,,
\quad\hbox{with}\quad
u(x)=\int_{-\infty}^\infty G(x,y)\,m(y)\,dy
\,,
\label{FH-beqn}
\end{equation}
may also be computed analytically for any value of $b$ and for any
symmetric Green's function. When $G(x,y)=G(y,x)$ and $b=2$, total momentum
is conserved. However, the $b-$pulson dynamics is only canonical when
$b=2$. For more details about the solution dependence on the parameter
$b$, see Holm and Staley [2003a,2003b],
\cite{HolmStaley[2003a]}, \cite{HolmStaley[2003b]}
and Degasperis, Holm and Hone [2002] \cite{DegasperisHolmHone[2002]}.
}

\section{Measure-valued momentum maps and solutions of geodesic flow in $n$
dimensions}

\subsection*{Measure-valued solution ansatz}

Based on the peakon solutions for the Camassa-Holm equation
\cite{CamassaHolm[1993]} and its generalizations to include the other
traveling-wave pulson shapes \cite{FringerHolm[2001]}, Holm \& Staley
\cite{HolmStaley[2003a]} introduced the following measure-valued ansatz
for the solutions of the vector EP equation (\ref{VectorEPeqn}),
\begin{equation}
\mathbf{m}(\mathbf{x},t)
=
\sum_{a=1}^N\int_s\mathbf{P}^a(s,t)\,
\delta\big(\,\mathbf{x}-\mathbf{Q}^a(s,t)\,\big)ds
\,,\quad
\mathbf{m}\in\mathbb{R}^n,\
s\in\mathbb{R}^k
\,,
\label{m-ansatz-intro}
\end{equation}
where the dimensions satisfy $k<n$. The fluid velocity
corresponding to the momentum solution ansatz (\ref{m-ansatz-intro}) is
given by
\begin{equation}
\mathbf{u}(\mathbf{x},t)
=
G*\mathbf{m}
=
\sum_{b=1}^N\int_{s'}\mathbf{P}^b(s^{\prime},t)\,
G\big(\,\mathbf{x},\mathbf{Q}^b(s^{\prime},t)\,\big)ds^{\prime}
\,,\quad
\mathbf{u}\in{\mathbb{R}^n}
\,,
\label{u-ansatz-intro}
\end{equation}
where $G(\mathbf{x},\mathbf{y})$ is the Green's function for the Helmholtz
operator in $n$ dimensions. These solutions are vector-valued functions
whose momenta are supported in
$\mathbb{R}^n$ on a set of $N$ surfaces (or curves) of  codimension
$(n-k)$ for $s\in\mathbb{R}^{k}$ with $k<n$.
In three dimensions, for example, they may be
supported on sets of points (vector peakons, $k=0$), quasi one-dimensional
filaments (strings, $k=1$), or quasi two-dimensional surfaces (sheets,
$k=2$). Substitution of the solution ansatz
(\ref{m-ansatz-intro}) into the EP equation (\ref{VectorEPeqn})
implies the following  integro-partial-differential equations (IPDEs)
for the evolution of such strings, or sheets,
\begin{eqnarray}
\frac{\partial }{\partial t}\mathbf{{Q}}^a (s,t)
\!\!&=&\!\!
\!\!\sum_{b=1}^{N} \int_{s'}\mathbf{P}^b(s^{\prime},t)\,
G(\mathbf{Q}^a(s,t),\mathbf{Q}^b(s^{\prime},t)\,\big)ds^{\prime}
\,,\label{IntDiffEqn-Q}\\
\frac{\partial }{\partial t}\mathbf{{P}}^a (s,t)
\!\!&=&\!\!
-\,\!\!\sum_{b=1}^{N} \int_{s'}
\big(\mathbf{P}^a(s,t)\!\cdot\!\mathbf{P}^b(s^{\prime},t)\big)
\, \frac{\partial }{\partial \mathbf{Q}^a(s,t)}
G\big(\mathbf{Q}^a(s,t),\mathbf{Q}^b(s^{\prime},t)\big)\,ds^{\prime}
\,.
\nonumber
\end{eqnarray}
%
Importantly for the interpretation of these solutions given later in Holm
and Marsden [2003] \cite{HolmMarsden[2003]}, the independent variables
$s\in\mathbb{R}^{k}$ turn out to be Lagrangian coordinates.  When evaluated
along the curve $\mathbf{x}=\mathbf{Q}^a(s,t)$, the fluid velocity
(\ref{u-ansatz-intro}) satisfies,
\begin{equation}\label{Qdot-ansatz-intro}
\mathbf{u}(\mathbf{x},t)\Big|_{\mathbf{x}=\mathbf{Q}^a(s,t)}
=
\sum_{b=1}^N\int_{s'}\mathbf{P}^b(s^{\prime},t)\,
G\big(\,\mathbf{Q}^a(s,t)
,\mathbf{Q}^b(s^{\prime},t)\,\big)ds^{\prime}
=
\frac{\partial\mathbf{Q}^a(s,t)}{\partial t}
\,.
\end{equation}
Consequently, the lower-dimensional support sets (defined on
$\mathbf{x}=\mathbf{Q}^a(s,t)$ and parameterized by coordinates
$s\in\mathbb{R}^{k}$) move with the fluid velocity. Moreover, equations
(\ref{IntDiffEqn-Q}) for the evolution of these support sets
are canonical Hamiltonian equations,
\begin{equation} \label{IntDiffEqns-Ham}
\frac{\partial }{\partial t}\mathbf{{Q}}^a (s,t)
=
\frac{\delta H_N}{\delta \mathbf{P}^a}
\,,\qquad
\frac{\partial }{\partial t}\mathbf{{P}}^a (s,t)
=
-\,\frac{\delta H_N}{\delta \mathbf{Q}^a}
\,.
\end{equation}
The corresponding Hamiltonian function $H_N:(\mathbb{R}^n\times
\mathbb{R}^n)^{\otimes N}\to \mathbb{R}$ is,
\begin{equation} \label{H_N-def}
H_N = \frac{1}{2}\!\int_s\!\int_{s'}\!\!\sum_{a\,,\,b=1}^{N}
\big(\mathbf{P}^a(s,t)\cdot\mathbf{P}^b(s^{\prime},t)\big)
\,G\big(\mathbf{Q}^a(s,t),\mathbf{Q}^b(s^{\prime},t)\big)
\,ds\,ds^{\prime}
\,.
\end{equation}
This is the Hamiltonian for canonical geodesic motion on the
cotangent bundle of a set of $N$ curves $\mathbf{Q}^a(s,t)$,
$a=1,\dots,N$, with respect to the metric given by $G$.

Hamiltonian geodesic dynamics under the measure-valued vector
solution ansatz (\ref{m-ansatz-intro}) for the EP equation
(\ref{VectorEPeqn}) was investigated numerically in Holm \& Staley [2003b]
\cite{HolmStaley[2003b]}. We refer to that paper for more details of
the solution dynamics.

The non-locality of the dynamics of the measured-valued solutions,
described in the literature, makes analytical progress difficult. In this paper,
we derive measured-valued solutions that are
rotationally symmetric. For this circular symmetry, the non-locality
integrates out and the motion reduces to a set of ordinary differential
equations. Most of the paper is devoted to these circular solutions, which
we call rotating peakons.  The set of solutions obeying
translational symmetry, but having
two velocity components, is studied in the appendix.
For a solution of $N$ planar peakons, there are $2N$ degrees of freedom,
with $2N$ positions and $2N$ canonically conjugate momenta. Hence, evolution of the $N$
planar peakon solution is governed by a set of $4N$ nonlocal partial
differential equations. These reduce to ordinary differential equations in
the presence of either rotational, or translational symmetry. We also
demonstrate that these solutions emerge from any initial condition with this
symmetry in the plane.
\subsubsection*{Potential applications of measure-valued solutions of
geodesic flow}

One of the potential applications of the two-dimensional version of this
problem involves the internal waves on the interface between two
layers of different density in the ocean.
\begin{figure}
\centering
\includegraphics[scale=0.5]{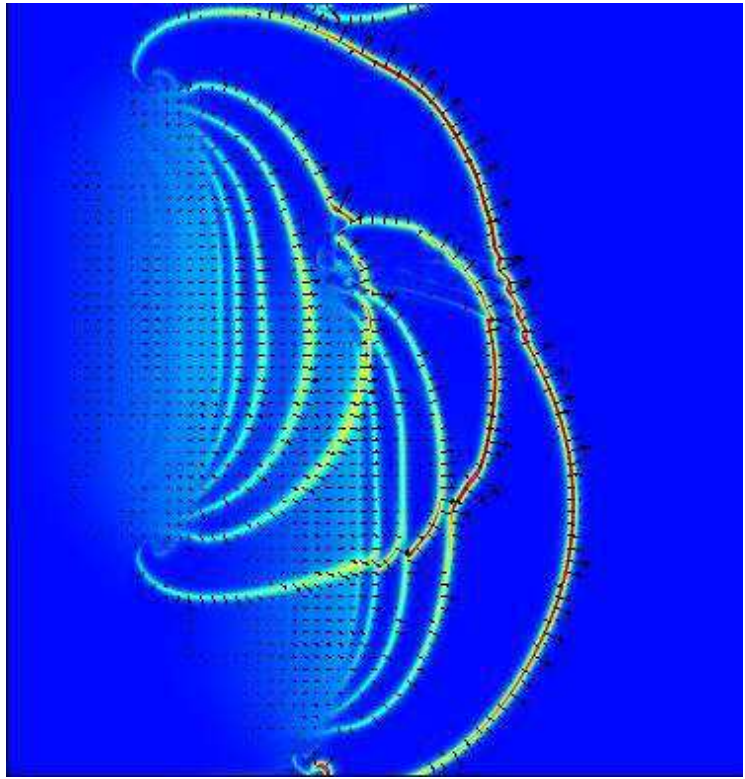}
\includegraphics[scale=0.3,angle=180]{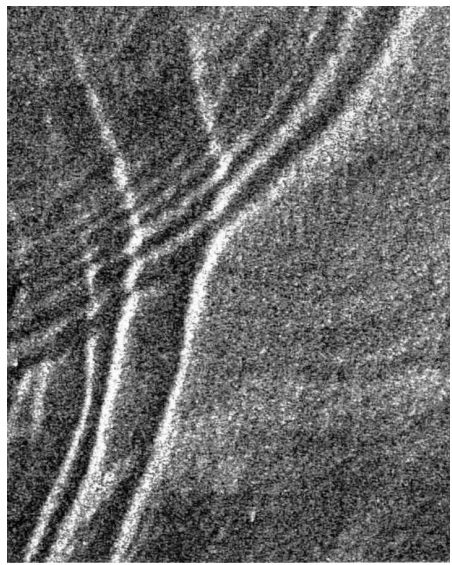}
\caption{Simulation of the full EP equation (\ref{VectorEPeqn}), courtesy of Martin
Staley (top). Internal waves in the South China Sea (bottom).
}
\label{iw-martin}
\end{figure}
Fig.~\ref{iw-martin} shows a striking agreement between two internal wave
trains propagating at the interface of different density levels in
the South China Sea, and the solution appearing in the simulations of
the EP equation (\ref{VectorEPeqn}) in two dimensions.
Inspired by this figure, we shall construct a theory of
propagating one-dimensional momentum filaments in two dimensions.
For other work on the 2D CH equation in the context of shallow water waves,
see Kruse, et al. (2001) \cite{Kruse-etal[2001]}.

Another potential application of the two-dimensional version of this
problem occurs in image processing for computational anatomy, e.g., brain
mapping from PET scans. For this application, one envisions the geodesic
motion as an optimization problem whose solution maps one measured
two-dimensional PET scan to another, by interpolation in three dimensions
along a geodesic path between them in the space of diffeomorphisms. In this
situation, the measure-valued solutions  of geodesic flow studied here
correspond to ``cartoon'' outlines of PET scan images. The geodesic
``evolution'' in the space between them provides a three dimensional image
that is optimal for the chosen norm.  For a review of this imaging
approach, which is called ``template matching'' in computational anatomy,
see Miller and Younes [2002]
\cite{MillerYounes[2001]}.


\subsection*{Peakon Momentum Map $J:\,T^*S \longrightarrow \mathfrak{g}^*$
in $n$ dimensions}

Holm and Marsden [2003] \cite{HolmMarsden[2003]} explained an important
component of the general theory underlying the remarkable reduced
solutions of the vector EP equation (\ref{VectorEPeqn}). In particular,
Holm and Marsden [2003] \cite{HolmMarsden[2003]} showed that the solution
ansatz (\ref{m-ansatz-intro}) for the momentum vector in the EP equation
(\ref{VectorEPeqn}) introduced in Holm and Staley [2003]
\cite{HolmStaley[2003b]} defines a {\bf momentum map} for the action of
diffeomorphisms on the support sets $\mathcal{S}$ of the Dirac delta
functions. These support sets are points on the real line for the CH
shallow water equation in one dimension. They are points, curves, or
surfaces in $\mathbb{R}^n$ for the vector EP equation (\ref{VectorEPeqn}) in
$n-$dimensions.

\paragraph{Momentum map definition.}
Let a group $G$ act on a manifold $S$, and lift the action of $G$ to the
cotangent bundle $T^*S$. A momentum map $J$ is a Poisson map from $T^*S$ to
$\mathfrak{g}^*$, the dual of the Lie algebra of $G$. (A map is Poisson,
provided it is coadjoint equivariant. In particular, $J$ maps the canonical
Poisson bracket on the image space $T^*S$ into the Lie-Poisson bracket on
the target space $\mathfrak{g}^*$.) In symbols, this is
\begin{eqnarray*}
\begin{array}{l}
J:\,(\mathbf{P},\mathbf{Q}) \in  T^*S
\longrightarrow \mathbf{m} \in \mathfrak{g}^*
\,,\\[3pt]
J:\, \{f,h\}_{can}(\mathbf{P},\mathbf{Q})
\longrightarrow \{f,h\}_{_{LP}}(\mathbf{m})
= \Big\langle \mathbf{m}\,,\,
\Big[\frac{\delta f}{\delta \mathbf{m}}\,,\,
\frac{\delta h}{\delta \mathbf{m}}\Big]
\Big\rangle
\,,\\[3pt]
\hbox{where}\quad
\big\langle \cdot\,,\, \cdot\big\rangle:
\mathfrak{g}^*\times\mathfrak{g}\longrightarrow\mathbb{R}
\,.
\end{array}
\end{eqnarray*}

The $n-$dimensional peakon momentum solution ansatz $J$ (for any
Hamiltonian) is given by Holm and Staley [2003] \cite{HolmStaley[2003b]}
as the superposition formula in (\ref{m-ansatz-intro}),
\begin{eqnarray}
J:\,\mathbf{m}(\mathbf{x},t)
=
\sum_{a=1}^N\int_s\mathbf{P}^a(s,t)\,
\delta\big(\,\mathbf{x}-\mathbf{Q}^a(s,t)\,\big)ds
\,,\quad
\mathbf{m}\in\mathbb{R}^n,\
s\in\mathbb{R}^k
\,.
\label{momentmap}
\end{eqnarray}
By direct substitution using the canonical $\mathbf{Q},\mathbf{P}$
Poisson brackets, one computes the Poisson property of the map $J$ in
$n$ Cartesian dimensions. Namely,
\begin{eqnarray}
\big\{m_i(\mathbf{x}),m_j(\mathbf{y})\big\}_{_{can}}
(\mathbf{P},\mathbf{Q})
=
-\,
\Big(
\frac{\partial}{\partial x\,^j}\,m_i(\mathbf{x})
+
m_j(\mathbf{x})\frac{\partial}{\partial x^i}
\Big)
\delta(\mathbf{x}-\mathbf{y})
\,,
\end{eqnarray}
in the sense of distributions integrated against a pair of smooth functions
of $\mathbf{x}$ and $\mathbf{y}$. This expression defines the Lie-Poisson
bracket $\{\cdot\,,\,\cdot\}_{_{LP}}(\mathbf{m})$ defined on the dual Lie
algebra $\mathfrak{g}^*$, restricted to momentum filaments supported on
the $N$ curves $\mathbf{x}=\mathbf{Q}^a(s,t)$, where $a=1,2,\dots,N$. Its
calculation demonstrates the following.

\begin{theorem}\label{mom-map}(Holm and Marsden [2003])
The momentum solution ansatz (\ref{m-ansatz-intro}), for
measure-valued solutions of the vector EP equation
(\ref{VectorEPeqn}), is a momentum map.
\end{theorem}
The Poisson property of the momentum map $J$ in (\ref{momentmap}) is, of
course, independent of the choice of Hamiltonian. This independence
explains, for example, why the map extends from peakons of a particular
shape, to the pulsons of any shape studied in Fringer and Holm [2001]
\cite{FringerHolm[2001]}. The solution ansatz (\ref{m-ansatz-intro}) now
rewritten as the momentum map
$J$ in (\ref{momentmap}) is also a Lagrange-to-Euler map, because the
momentum is supported on filaments that {\it move with the fluid velocity}.
Hence, the motion governed by the vector EP equation (\ref{VectorEPeqn})
occurs by the action of the diffeomorphisms in $G$ on the support set of
the fluid momentum, whose position and {\it canonical} momentum are defined
on the cotangent bundle $T^*S$ of the space of curves $S$. This
observation informs the study of geodesic motion governed by equation
(\ref{VectorEPeqn}). For complete details and definitions, see  Holm and
Marsden [2003]
\cite{HolmMarsden[2003]}.

\subsubsection*{Peakon momentum map
$J:\,T^*S \longrightarrow \mathfrak{g}^*$
on a Riemannian manifold}

The goal of the present work is to characterize the measure-valued
solutions of the vector EP equation (\ref{VectorEPeqn}), by using the
momentum map $J$ in (\ref{momentmap}) when $S$ is the space of concentric
circles in the plane. The motion and interactions of these measure-valued
solutions may be either purely radial (circles of peakons), or they may
also have an azimuthal component (rotating circles of peakons). To
accomplish this goal, we employ a result from  Holm and Staley [2003b]
\cite{HolmStaley[2003b]} that on a Riemannian manifold $M$ with metric
determinant $\det{g}(\mathbf{x})$, the measure-valued momentum ansatz
(\ref{m-ansatz-intro}) becomes
\begin{equation}
\mathbf{m}(\mathbf{x},t)
=
\sum_{a=1}^N\int_s \mathbf{P}^a(s,t)\,
\frac{\delta\big(\,\mathbf{x}-\mathbf{Q}^a(s,t)\,\big)}
{\sqrt{\det{g}}}\,ds
\,,\quad
\mathbf{m}\in{M},\
s\in\mathbb{R}^k
\,.
\label{m-ansatz-Riemannian}
\end{equation}
This solution ansatz is also a momentum map, as shown in Holm and
Marsden [2003] \cite{HolmMarsden[2003]}. On a Riemannian manifold, the
corresponding Lie-Poisson bracket for the momentum on its support set
becomes
\begin{eqnarray}\label{momentum-map-bracket1}
\big\{m_i(\mathbf{x})\,,\,m_j(\mathbf{y})\big\}
=
-\,
\Big(
m_j(\mathbf{x})\frac{\partial}{\partial x^i}
+
\frac{1}{\sqrt{\det{g}}}\,
\frac{\partial}{\partial x\,^j}\,
\sqrt{\det{g}}\,m_i(\mathbf{x})\Big)
\frac{\delta(\mathbf{x}-\mathbf{y})}{\sqrt{\det{g}}}
\,.
\label{LP-bracket}
\end{eqnarray}
For example, in cylindrical  symmetry one has
$\sqrt{\det{g}}=r$
and the vector $\mathbf{m}$ depends
only on the radial coordinate $r$. For solutions with these symmetries, the
Lagrangian label coordinate $s$ is unnecessary, as we shall see in the
cylindrical case, and the equations for
$\mathbf{Q}^a$ and $\mathbf{P}^a$ will reduce to ordinary differential
equations in time.

\section{Lie-Poisson bracket for rotating concentric
circles of peakons}

\subsubsection*{Azimuthal relabeling symmetry for rotating circular peakons}

We shall consider the dynamics of circles of peakons, whose
motion may have both radial and azimuthal components. These are
rotating circular peakons. Suppose one were to mark a Lagrangian point on
the $a-$th circle, $a=1,\dots,N$. Then the change in its azimuthal angle
$\phi_a(t)$ could be measured as it moved with the azimuthal fluid velocity
$u_\phi$ along the  the $a-$th circle as its radius $r=q_a(t)$ evolved.
Translations in the Lagrangian azimuthal coordinate would shift the mark,
but this shift of a Lagrangian label would have no effect on the Eulerian
velocity dynamics of the system. Such a Lagrangian relabeling would be a
symmetry for any Hamiltonian depending only on Eulerian velocity. Thus, the
azimuthal relabeling would result in the conservation of its canonically
conjugate angular momentum $M_a$, which generates the rotation
corresponding to the relabeling symmetry of the $a-$th circle. The
$a-$th circle would be characterized in phase space by its radius
$r=q_a(t)$, and its canonically conjugate radial momentum, denoted as
$p_a$. The rotational degree of freedom of the $a-$th circle would be
represented by its conserved angular momentum $M_a$ and its ignorable
canonical azimuthal angle $\phi_a$. The only nonzero canonical Poisson
brackets among these variables are,
\begin{equation}
\{q_a,p_b\}_{can}=\delta_{ab}
\quad\hbox{and}\quad
\{\phi_a,M_b\}_{can}=\delta_{ab}
\,.
\label{can-spin-brac}
\end{equation}

\subsubsection*{Momentum map for rotating circular peakons}

In terms of their $4N$ canonical phase space variables
$(q_a,p_a,\phi_a,M_a)$, with $a=1,2,\dots,N$, the superposition formula
(\ref{momentmap}) for
$N$ rotating circular peakons may be expressed as,
\begin{equation}
\mathcal{J}\!\!:\,
\mathbf{m}(r,t)
=
\sum_{a=1}^N \big(\,
p_a(t)\,\mathbf{\hat{r}}
+
\frac{M_a}{q_a(t)}\,\boldsymbol{\hat{\phi}}
\,\big)
\frac{\delta\big(\,r-q_a(t)\,\big)}
{r}
\,.
\label{spin-peakon-momentmap}
\end{equation}
We shall first verify that this formula is a momentum map, and then in
section \ref{eqn-der} we shall derive it, by requiring it to be a valid
solution ansatz for the geodesic EP equation (\ref{VectorEPeqn}) in polar
coordinates. As a consequence, the motion governed by the system of partial
differential equations (\ref{mr-intro}) and (\ref{mphi-intro}) for geodesic
motion in the plane with azimuthal symmetry has a finite dimensional
invariant manifold in the $2N-$dimensional canonical phase space
$(q_a,p_a)$ for each choice of the $N$ angular momentum values $M_a$, with
$a=1,2,\dots,N$. Later, we shall also examine numerical studies of these
solutions when the kinetic energy is chosen to be the $H^1$ norm of the
azimuthally symmetric fluid velocity.

By direct substitution using the canonical  Poisson brackets in
(\ref{can-spin-brac}), one computes the Poisson property of the map
$\mathcal{J}$ in (\ref{spin-peakon-momentmap}). Namely,
\begin{eqnarray}
\big\{m_r(r),m_r(r')\big\}_{can}%
(\mathbf{p},\mathbf{q})
&=&
-\,
\Big(
\frac{1}{r}\frac{\partial}{\partial r} r m_r(r)
+ m_r(r)\frac{\partial}{\partial r}
\Big)
\frac{\delta(r-r')}{r}
\,,\nonumber\\
\big\{m_r(r),r ' m_\phi(r')\big\}_{can}%
(\mathbf{p},\mathbf{q})
&=&
-\,
r m_\phi(r) \frac{\partial}{\partial r}\,
\frac{\delta(r-r')}{r}
\,,\nonumber\\
\big\{rm_\phi(r),m_r(r')\big\}_{can}%
(\mathbf{p},\mathbf{q})
&=&
-\,
\frac{1}{r}\frac{\partial}{\partial r} r^2 m_\phi(r)
\,
\frac{\delta(r-r')}{r}
\,,\nonumber\\
\big\{rm_\phi(r),rm_\phi(r')\big\}_{can}%
(\mathbf{p},\mathbf{q})
&=&
0
\,.
\label{LPB-cyl}
\end{eqnarray}
These equalities are written in the sense of distributions integrated
against a pair of smooth functions of $r$ and $r'$. They demonstrate the
Poisson property of the map $\mathcal{J}$ in (\ref{spin-peakon-momentmap}),
which is also the solution ansatz for the rotating circular peakons. They
also express the Lie-Poisson bracket
$\{\cdot\,,\,\cdot\}_{_{LP}}(m_r,rm_\phi)$ for momentum filaments defined
on the dual Lie algebra $\mathfrak{g}^*$ and restricted to the support set
of these solutions. Hence, we have demonstrated the following:

\begin{proposition}\label{mom-map-spin-peakon}
The map $\mathcal{J}$ in (\ref{spin-peakon-momentmap}) is a momentum map.
\end{proposition}

On comparing the formulas in (\ref{LPB-cyl}) with the Hamiltonian operator
$\mathcal{D}$ for the continuous solutions in (\ref{Ham-op}), one sees that
the momentum map (\ref{spin-peakon-momentmap}) essentially restricts the
Lie-Poisson bracket with Hamiltonian operator $\mathcal{D}$ to its support
set. Next, we shall derive the momentum map (\ref{spin-peakon-momentmap}) by
requiring it to be a valid solution ansatz for the geodesic EP equation
(\ref{VectorEPeqn}) in polar coordinates.

\rem{
\fbox{\color{red}I suggest we replace this paragraph with the blue one
below -- DH}

As we shall see below, the evolution
of these momentum filaments will be described by one-dimensional non-local
Partial Differential Equations (PDEs). In general, these PDEs are
difficult to analyze, so in order to make useful progress, we proceed with
the detailed analysis of some particular cases, when the non-locality can
be integrated out and the motion of the momentum filaments can be
described by a set of Ordinary Differential Equations (ODEs).
\comment{VP:
I love the Schlichting way of putting things!

\fbox{\color{red}I'm not so sure! I would put it as in the next (blue)
comment -- DH}}

\comment{\color{blue}
The goal of the present work is to use the Lagrange-to-Euler momentum map
(\ref{m-ansatz-intro}) to characterize the solutions of the vector EP
equation with cylindrical symmetry. To accomplish this, we notice that on
a Riemannian manifold $M$ with metric determinant
$\det{g}(\mathbf{x})$, the measure-valued momentum ansatz
(\ref{m-ansatz-intro}) becomes
\begin{equation}
\mathbf{m}(\mathbf{x},t)
=
\sum_{a=1}^N\int_s \mathbf{P}^a(s,t)\,
\frac{\delta\big(\,\mathbf{x}-\mathbf{Q}^a(s,t)\,\big)}
{\det{g}(\mathbf{x})}\,ds
\,,\quad
\mathbf{m}\in{M}
\,.
\label{m-ansatz-Riemannian}
\end{equation}
For example, in cylindrical and spherical symmetry
$\det{g}(\mathbf{x})=r,\,r^2$, respectively, and the vector
$\mathbf{m}$ has only a radial component. In these one-dimensional
cases, the Lagrangian label coordinate $s$ is unnecessary and the
equations again reduce to ordinary differential equations.\\
{\color{red}This is enough for writing the momentum map in
cylindrical and spherical coordinates. -- DH} }

\section{Peakon filaments on the plane\\
\fbox{\color{red}I suggest we drop this section -- DH}}

Suppose we consider a generalized solution of the equation
(\ref{VectorEPeqn}) in the form
\beq
\mathbf{m}=\sum_i\int_{s_i}{\bf p}(t,s_i) \delta({\bf r}-{\bf q_i}(t,s_i)).
\label{filament}
\eeq
Such solution describes a set of momentum filaments of strength ${\bf
p_i}$ concentrated
on the lines ${\bf r=q_i}$, and $s_i$ is the arc length along the $i$-th
filament.
It can be proven that the motion of filaments is canonical with the
Hamiltonian being equal to
\beq    H=(1/2) \sum_{i,j} \int_{s_i} \int_{s_j}
       (\mathbf{P}_i(s_i,t){\cdot}\mathbf{P}_j(s_j,t))
                  G(q_i(s_i,t),q_j(s_j,t))
\eeq
with $\mathbf{q}_i(s_i,t)$ and $\mathbf{p}_i(s_i,t)$ are canonically
conjugate pairs of vectors in two dimensions.
The equation of motion for the coordinates ${\bf q_i}(s_i,t)$ and
$\mathbf{p}_i(s_i,t)$ are
\beq
\mathbf{p}_i(s_i,t)=-\frac{\partial H}{\partial \mathbf{q}_i}
\eeq
\beq
\mathbf{q}_i(s_i,t)=\frac{\partial H}{\partial \mathbf{p}_i}
\eeq

In general case, these evolution equations for the position of the filament
are non-local as they involve the integration along the filaments.
Thus, very little analytical progress can be made in general case.
However, we will show
how these equations can be integrated exactly for two particular cases
  of filaments being parallel lines or concentric rings on a plane.
}

\section{Azimuthally symmetric peakons}\label{eqn-der}
\subsection{Derivation of equations}

We seek azimuthally symmetric solutions of the geodesic EP equation
(\ref{VectorEPeqn}) in polar coordinates
$(r,\phi)$, for which
\beq
\mathbf{m}=m_r(r,t)\mathbf{\hat{r}}+m_\phi(r,t)\boldsymbol{\hat{ \phi}}
\equiv
(m_r(r,t),m_\phi(r,t))
\,.
\label{cyl-mom}
\eeq
We shall derive the momentum map (\ref{spin-peakon-momentmap}) and the
canonical Hamiltonian equations for its parameters $(q_a,p_a,M_a)$
by assuming solutions in the form,
\beq
\mathbf{m}(r,t)=\sum_{i=1}^N \left( p_i(t) \mathbf{\hat{r}}
+ v_i(t) \boldsymbol{\hat{ \phi}} \right) \frac{\delta(r-q_i(t))}{r}
\,.
\label{mvcyl}
\eeq
These solutions represent concentric cylindrical momentum filaments which
are rotating around the origin. The corresponding velocity components are
obtained from
\beq
  \big( u_r(r,t), u_\phi(r,t)\big)=\int r' G(r,r')
\big( m_r(r',t), m_\phi(r',t)\big)
\,\mbox{d} r'
\,, \label{ur}
  \eeq
where $G(r,r')=G(r',r)$ is the (symmetric) Green's function for the radial
Helmholtz operator given in formula (\ref{greensfcn}). Hence, the fluid
velocity corresponding to the solution ansatz (\ref{mvcyl}) assumes the
form,
\beq
\mathbf{u}(r,t)=\sum_{j=1}^N
\left( p_j(t) \mathbf{\hat{r}}
+ v_j(t) \boldsymbol{\hat{ \phi}} \right)G(r,q_j(t))
\,,
\label{vel-cyl}
\eeq
with Green's function $G(r,q_j(t)) $ as in formula (\ref{greensfcn}).
In addition, the kinetic energy of the system is
given by
\beq
K\!E({\bf p}, {\bf q}, {\bf v})
=
\frac{1}{2}\int \mathbf{u} \cdot \mathbf{m}\, r \mbox{d} r
=
\frac{1}{2} \sum_{i,j=1}^N \Big( p_i p_j  + v_i v_j\Big) G(q_i,q_j)
\,.
\label{ergcyl}
\eeq
%
\rem{
We shall verify the solution ansatz (\ref{mvcyl}) directly.
This requires writing
the vector Helmholtz operator (\ref{CHmom}) acting on
$r$- and $\phi$- components of velocity $\mathbf{u}$. Without angular
dependence, the Helmholtz operator yields
\[
m_r=u_r-\left(\frac{\partial^2 u_r}{\partial r^2}
+ \frac{1}{r} \frac{\partial u_r}{\partial r} -\frac{u_r}{r^2} \right)
\,,\quad
m_\phi=u_\phi-\left(\frac{\partial^2 u_\phi}{\partial r^2} + \frac{1}{r}
\frac{\partial u_\phi}{\partial r} - \frac{u_\phi}{r^2} \right)
\]
Consequently, the Green's function for both $r$ and $\phi$ components

Given the radial momentum $m(r,t)$, the corresponding radial
velocity $u(r,t)$ is given by the convolution $u(r)=G \star
m=\int_0^r G(r,\xi) m(\xi)
\xi
\mbox{d} \xi$ (an extra factor $\xi$ arises in
cylindrical geometry) with the Green's function $G(r,\xi)$
\beq
G(r,\xi)=\left\{ \begin{array}{cc}
I_1(\xi) K_1(r) \quad\hbox{for}\quad & \xi< r\,, \\
I_1(r) K_1( \xi) \quad\hbox{for}\quad & r< \xi\,,
\end{array}
\right.
\label{g}
\eeq
where $I_1$ and $K_1$ are Modified Bessel's functions.

is
still given by  (\ref{g}).
The corresponding velocity is given by
\beq
  \big( u_r(r,t), u_\phi(r,t)\big)=\int r' G(r,r')
\big( m_r(r',t), m_\phi(r',t)\big)
\mbox{d} r'
  \label{ur}
  \eeq
The $r$ and $\phi$ components of $ (\nabla \mathbf{u})^T \cdot\mathbf{m}$
have the following forms (assuming no $\phi$-dependence):
  \beq
\left(  (\nabla \mathbf{u})^T \cdot\mathbf{m} \right)_r
=
m_\phi(r) u\,'_\phi(r)+m_r(r) u\,'_r(r)
  \eeq
\beq
\left(  (\nabla \mathbf{u})^T \cdot\mathbf{m} \right)_\phi= \frac{m_r(r)
u_\phi(r)-m_\phi(r) u_r(r)}{r}
  \eeq
Consequently, the momentum equation (\ref{VectorEPeqn}) has
$r$-component:
\beq
\frac{\partial m_r}{\partial t}+ u_r \frac{\partial m_r}{\partial r} + 2
m_r \frac{\partial u_r}{\partial r } +
\frac{m_r u_r - m_\phi u_\phi}{r}
+ m_\phi \frac{\partial
u_\phi}{\partial r}=0,
\label{mr1}
\eeq
and  $\phi$ component:
\beq
\frac{\partial m_\phi}{\partial t}+ u_r \frac{\partial m_\phi}{\partial
r} + m_\phi \frac{\partial u_r}{\partial r } +
\frac{2 m_\phi u_r}{r} =0.
\label{mphi1}
\eeq
}
Substitution of the solution ansatz (\ref{mvcyl}) for the momentum
and its corresponding velocity (\ref{vel-cyl}) into the radial equation
(\ref{mr-intro}) gives the system,
\begin{eqnarray*}
&&\hspace{-5mm}
\sum_i \left( \dot{p}_i \frac{\delta(r- q_i)}{r}- p_i
\dot{q}_i
\frac{\delta\,'(r- q_i)}{r} \right)
\nonumber\\&&\hspace{-5mm}+\
  \sum_{i,j} \Bigg\{
  p_i p_j G(r,q_j)
\left[-\frac{\delta(r-
q_i)}{r^2}+
\frac{\delta\,'(r- q_i)}{r} \right] + 2 p_i p_j \frac{\delta(r-
q_i)}{r} \frac{\partial G}{\partial r}(r,q_j)
\nonumber\\&&\hspace{1cm}+\
(p_i p_j-v_i v_j) \frac{\delta(r- q_i)}{r^2} G(r,q_j)
+ v_i v_j \frac{\delta(r-
q_i)}{r} \frac{\partial G}{\partial r}(r,q_j)
\Bigg\}
=0
\,.
\label{intmr}
\end{eqnarray*}
Multiplying this system by the smooth test function $r\psi(r)$ and
integrating with respect to $r$ yields dynamical equations for $p_i$ and
$q_i$. In particular, the
$\psi(q_i)$ terms yield
\beq
\dot{p}_i=- \sum_j  \left( p_i p_j \frac{\partial G}{\partial q_i}
+ v_i v_j
\left\{ \frac{\partial G(q_i,q_j)}{\partial q_i} -
\frac{G(q_i,q_j)} {q_i} \right\} \right)
\,,
\label{peqrphi}
\eeq
and, after integrating by parts, the $\psi'(q_i)$ terms yield
\beq
\dot{q}_i= \sum_j p_j G(q_i,q_j)
\,.
\label{qeqrphi}
\eeq
By equation (\ref{vel-cyl}) we see that
$\dot{q}_i(t)=\mathbf{\hat{r}}\cdot\mathbf{u}(q_i,t)$, so the radius
of the $i-$th cylinder moves with the radial velocity of the flow.

This procedure is repeated for the $\phi$ component of the EP  equation, by
substituting the solution ansatz (\ref{mvcyl},\ref{vel-cyl}) into
equation (\ref{mphi-intro}), to find the system
\begin{eqnarray*}
&&\sum_i
\left( \dot{v}_i \frac{\delta(r- q_i)}{r}
- v_i \dot{q}_i \frac{\delta\,'(r-q_i)}{r} \right)
\\&&+\
  \sum_{i,j} \Bigg\{ v_i p_j G(r,q_j) \left[-\frac{\delta(r- q_i)}{r^2}+
\frac{\delta\,'(r- q_i)}{r} \right]
\\&&\hspace{1cm}
+\, v_i p_j \frac{\delta(r- q_i)}{r} \left( \frac{\partial G}{\partial
r}(r,q_j) + 2\frac{G(r,q_j)}{r} \right) \Bigg\}=0
\,.
\label{intmphi}
\end{eqnarray*}
Upon multiplying this system by $r \psi(r)$ and integrating with
respect to $r$, the term proportional to
$\psi'(q_i)$ again recovers exactly the $q_i-$equation (\ref{qeqrphi}). The
term proportional to
$\psi(q_i)$ gives
\beq
\dot{v}_i=-\,\frac{v_i}{q_i}\sum_j  p_j G(q_i,q_j)
=
-\,v_i\frac{\dot{q}_i}{q_i}
\,,
\label{veqcyl}
\eeq
after using the $q_i-$equation (\ref{qeqrphi}) in the last step.
This integrates to
\beq
v_i q_i = M_i = const,
\label{angmom}
\eeq
where $M_i$ are $N$ integration constants.
  From the Hamiltonian viewpoint, this was expected: The angular
momentum $M_i$ is conserved for each individual circular peakon, because
each circle may be rotated independently without changing the energy.

Equations (\ref{peqrphi}) and (\ref{qeqrphi}) for $p_i$ and $q_i$
may now be recognized as Hamilton's canonical equations with
Hamiltonian,
\beq
H({\bf p}, {\bf q}, {\bf M})
=
\frac{1}{2}\!\!\int\!\!\mathbf{u} \cdot \mathbf{m}\, r \mbox{d} r
=
\frac{1}{2} \sum_{i,j=1}^N \Big( p_i p_j
+ \frac{M_i M_j}{q_i q_j}\Big) G(q_i,q_j)
\,.
\label{hamcyl}
\eeq
This is the same Hamiltonian as obtained from substituting the momentum map
(\ref{spin-peakon-momentmap}) into the kinetic energy $K\!E$ in equation
(\ref{ergcyl}).
%
\rem{
The Hamitonian equations for the parameters
$(\mathbf{p}, \mathbf{q}, \boldsymbol{\phi}, \mathbf{M})$ are obtained
from the canonical Poisson brackets, in which
\beq
\{q_i,p_j\}=\delta_{ij}
\,,\quad
\{\phi_i,M_j\}=\delta_{ij}
\,,
\label{canonPB}
\eeq
and the other Poisson brackets vanish.
}
Hence, we may recover the reduced equations (\ref{peqrphi}) for $p_i$,
(\ref{qeqrphi}) for $q_i$ and (\ref{angmom}) for $M_i$ from the
Hamiltonian (\ref{hamcyl}) and the canonical equations,
\beq
\label{caneqs}
\dot{p}_i
=
-\,\frac{\partial H}{\partial q_i}
\,,\quad
\dot{q}_i
=
\frac{\partial H}{\partial p_i}
\,,\quad
\dot{M}_i
=
-\,\frac{\partial H}{\partial \phi_i}
=0
\,.
\eeq
This result proves the following:

\begin{proposition}\label{mom-map-as-solm}
The momentum map (\ref{spin-peakon-momentmap})
is a valid ansatz for rotating peakon solutions
of the Euler-Poincar\'e
equation (\ref{VectorEPeqn}) for geodesic motion.
\end{proposition}

\subsection{Solution properties}

The remaining canonical equation for the $i-$th Lagrangian angular
frequency is,
\beq
\dot{\phi}_i
=
\frac{\partial H}{\partial M_i}
=
  \sum_j \frac{M_j}{q_iq_j} G(q_i,q_j)
=
\frac{ 1}{q_i}\sum_j v_j(t) G(q_i,q_j)
\,.
\label{canonPB}
\eeq
Thus, as expected, the ignorable canonical angle variables
$\boldsymbol{\phi}=\{\phi_i\}$, with $i=1,2,\dots,N$, decouple from the
other Hamiltonian equations. In addition, we see that
\beq
q_i\dot{\phi}_i(t)=\boldsymbol{\hat{\phi}}\cdot\mathbf{u}(q_i,t)
\,,
\eeq
so the angular velocity of the $i-$th cylinder {\it also} matches the
angular velocity of the flow. Therefore, we have shown:

\begin{proposition}\label{LagEul-mom-map}
The canonical Hamiltonian parameters in the momentum map and solution
ansatz (\ref{spin-peakon-momentmap}) provide a Lagrangian description in
cylindrical symmetry of the flow governed by the Eulerian EP equation for
geodesic motion (\ref{VectorEPeqn}).
\end{proposition}

\paragraph{Angular momentum, fluid circulation and collapse to the center.}
Finally, the fluid circulation of the $i-$th concentric circle $c_i$,
which is traveling with velocity $\mathbf{u}$, may be
computed from equations (\ref{cyl-mom}) and (\ref{mvcyl}) (with a slight
abuse of notation) as,
\beq
\oint_{c_i(\mathbf{u})}\mathbf{m}\cdot\mbox{d}\mathbf{x}
=
\oint_{c_i(\mathbf{u})}rm_\phi\mbox{d}\phi
=
2\pi v_i=\frac{2\pi M_i}{q_i}.
\label{spin}
\eeq
We see that the ``angular velocity'' $v_i=M_i/q_i$ is the
fluid  circulation of the $i-$th concentric circle. Since the angular
momentum $M_i$ of the $i-$th circle is conserved, its
circulation $v_i(t)$ varies inversely with its radius. Consequently, this
circulation would diverge, if the $i-$th circle were to collapse to the
center with nonzero angular momentum.

\rem{
\subsection{Derivation of equations}
We seek radially symmetric solutions in polar coordinates $(r,\phi)$, for
which  $\mathbf{m}=(m(r,t),0)$, $\mathbf{u}=(u(r,t),0)$.  Then, $u$ and
$m$ satisfy:
\beq
m_t+ u m_r + 2 m u_r + m \frac{u}{r} =0
\label{radmeq}
\eeq
with subscripts $r,t$ denoting the partial derivatives with
respect to $r$ and $t$, respectively. For radial symmetry, the Helmholtz
relation between $m$ and $u$ becomes
\beq m=u-\frac{\partial^2 u}{\partial
r^2}- \frac{1}{r} \frac{\partial u}{\partial r}+\frac{u}{r^2}
\label{helmoltzr} \eeq
Given the radial momentum $m(r,t)$, the corresponding radial
velocity $u(r,t)$ is given by the convolution $u(r)=G \star
m=\int_0^r G(r,\xi) m(\xi)
\xi
\mbox{d} \xi$ (an extra factor $\xi$ arises in
cylindrical geometry) with the Green's function $G(r,\xi)$ \beq
G(r,\xi)=\left\{ \begin{array}{cc}
I_1(\xi) K_1(r), & \xi< r\,, \\
I_1(r) K_1( \xi), & r< \xi\,.
\end{array}
\right.
\label{g}
\eeq
Here $I_1$ and $K_1$ are Modified Bessel's functions of order $1$.  Unlike
its one-dimensional planar analog, the radial equation (\ref{radmeq}) is
not known to be integrable, in the sense of allowing reformulation as an
isospectral problem. We shall seek cylindrical peakon solutions in
the form of the momentum map,
\beq
m(r,t)=\sum_{i=1}^N p_i(t) \frac{\delta(r- q_i(t))}{r}
\,.
\label{mdelta}
\eeq
  To find equations of motion for $p_i(t)$ and $q_i(t)$, we
first express $u$ as $G \star m$ with $G$ given by (\ref{g}):
\beq
u(r,t)=G \star m = \int G(r,r')\, m(r',t)\,r'dr'
=\sumin p_i(t) G(r,q_i)
\,.\label{ueq}
\eeq
Let us now substitute (\ref{mdelta}) and (\ref{ueq}) into (\ref{radmeq})
and multiply
by an arbitrary test function $\phi(r)$ and integrate by parts as
necessary. We obtain, term by term:
\[
\int \phi m_t \,rdr
= \sumin \phi(q_i) \dot{p}_i + \phi'(q_i) p_i \dot{q}_i \]
\[
\int u m_r \phi \,rdr
=-\sumijn p_i p_j \phi'(q_i) G(q_i,q_j)
- \sumijn p_i p_j \phi(q_i)
  \frac{\partial G}{\partial r}(q_i,q_j)
+\frac{G(q_i,q_j)}{q_i}
\]
\[
\int m u_r \phi \,rdr
= \sumijn p_i p_j \phi(q_i) \frac{\partial G}{\partial r}(q_i,q_j)
\]
\[
\int m \frac{u}{r} \phi \,rdr
= \sumijn p_i p_j \phi(q_i)
\frac{G(q_i,q_j)}{q_i}
\]
Collecting terms multiplying $\phi(q_i), \phi'(q_i)$
gives {\it canonical} equations for $\dot{p}_i$, $\dot{q}_i$ respectively:
\beq
\dot{p}_i =- \sumjn p_i p_j \, \frac{\partial G}{\partial r}(q_i,q_j)
=
-\,\frac{\partial H}{\partial q_i}
\,,\qquad
\dot{q}_i = \sumjn p_j G(q_i,q_j).
=
\frac{\partial H}{\partial p_i}\label{peq}
\eeq
where the Hamiltonian $H$ is given by,
\beq
H = \half\sumijn p_i p_j G(q_i,q_j)
=
\half \int mu \,rdr
\eeq
The velocity equation (\ref{ueq}) implies
\beq
u(q_i,t)=\sumjn p_j(t) G(q_i,q_j) = \dot{q}_i(t)
\,.
\label{ueq-lag}
\eeq
Therefore, the position variable $q_i(t)$ moves with the velocity of
the flow -- that is, the $q_i$ are Lagrangian variables.
}

\section{Numerical Results for Radial Peakons}
\subsection{Radial peakon collisions}

\begin{figure}
\centering
\includegraphics[scale=0.5]{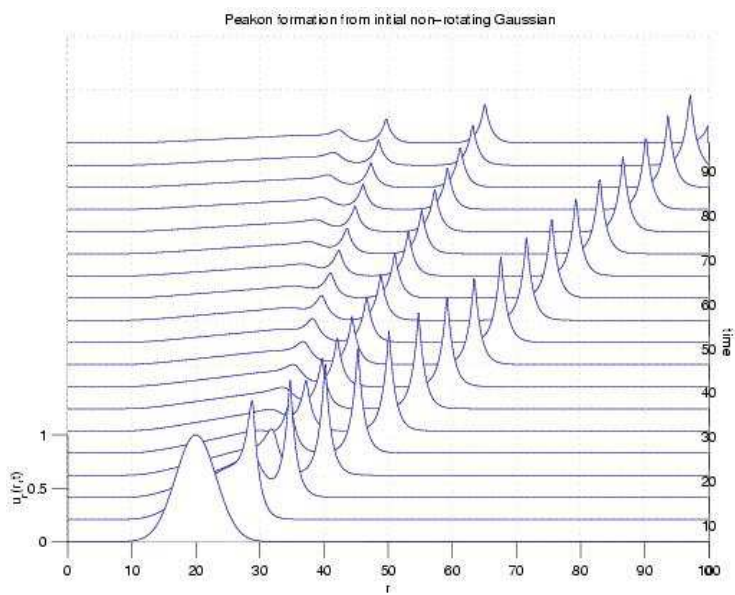}
\caption{The initial value problem: a Gaussian profile splits into
peakons.}
\label{IVP-rad}
\end{figure}

We consider purely radial solutions of equation (\ref{mr-intro}), with
$m_\phi=0$, which satisfy
\beqa
\frac{\partial m_r}{\partial t}
&=&
-\, \frac{1}{r}\partial_r\big(rm_ru_r\big)
- m_r\partial_ru_r
\,.
\label{radmeq}
\eeqa
Such radial solutions have no azimuthal velocity. Without azimuthal
velocity, the vector peakon solution ansatz (\ref{mvcyl}) for
momentum reduces to the scalar relation,
\beq
m_r(r,t)=\sum_{i=1}^N p_i(t) \frac{\delta(r-q_i(t))}{r}
\,.
\label{mv-rad}
\eeq
The corresponding radial velocity is
\beq
u_r(r,t)=\sum_{i=1}^N p_i(t)G(r,q_i(t))
\,,
\label{vel-rad}
\eeq
where the Green's function $G(r,q_i(t))$ for the radial Helmholtz
operator is given by formula (\ref{greensfcn}).  Radial peakons of this
form turn out to be the building blocks for the solution of any radially
symmetric initial value problem. We have found numerically that the initial
value problem for equation (\ref{mv-rad}) with any initially confined radial
distribution of velocity quickly splits up into radial peakons. This
behavior is illustrated in Fig.~\ref{IVP-rad}.
\rem{
The initial distribution of
velocity splits almost immediately into radial peakons moving inward and
outward that undergo collision interactions amongst each other. Those radial
peakons moving inward without obstruction by any others will collapse to
the center, and rebound reversibly.
}
The initial distribution of velocity splits almost immediately into
a train of radial peakons arranged by height, or,
equivalently, speed.

\begin{figure}
\centering
\includegraphics[scale=0.5]{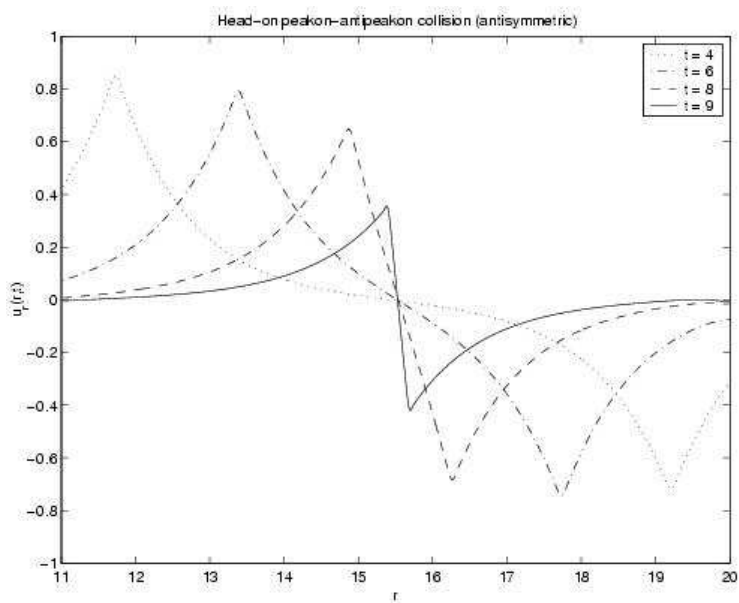}
\caption{Peakon-antipeakon collision of equal strength.}
\label{p-ap-equal}
\end{figure}

\begin{figure}
\centering
\includegraphics[scale=0.5]{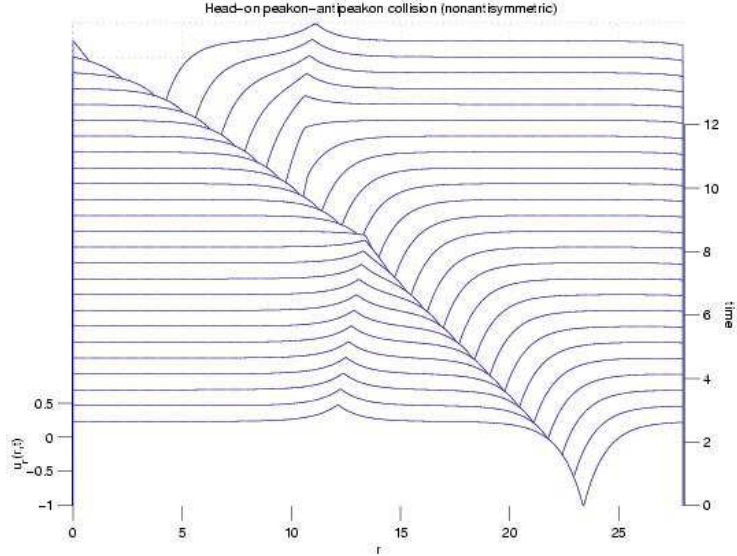}
\caption{Peakon-antipeakon collision of unequal strength. The
   smaller peakon's trajectory undergoes a large phase shift.}
\label{p-ap-neq}
\end{figure}

The head-on ``peakon-antipeakon'' collisions are of special interest. In
the case of equal strength radial peakon-antipeakon collisions, the solution
appears to develop infinite slope in finite time, see
Fig.~\ref{p-ap-equal}. This behavior is also known to occur for
peakon-antipeakon collisions on the real line.
If the strengths of the peakon and  antipeakon are not equal, then the
larger one of them seems to `plow' right through the smaller one.
This is shown in Fig.~\ref{p-ap-neq}.


The figures shown were produced from numerical simulations of the
Eulerian PDE (\ref{radmeq}).  The momentum $m_r$ was advanced in
time using a fourth-order Runge-Kutta method.  The time step was
chosen to ensure the Hamiltonian
$\frac{1}{2}\int\mathbf{m}\cdot\mathbf{u}\,r\,dr$ was conserved
to within 0.1\% of its initial value.  The spatial discretizations
ranged from $dr=10^{-4}$ to $dr=0.02$ depending on the desired
resolution and the length of the spatial domain,
and the spatial derivatives
were calculated using finite differences.
Fourth- and fifth-order
centered differencing schemes were used for the first and second
derivatives,
respectively.  The momentum $m_r$ was found from the velocity $u_r$
using the finite difference form of the radial Helmholtz operator,
and the velocity $u_r$ was found from the momentum $m_r$ by
inverting the radial Helmholtz matrix.  For the peakon interaction
simulations, the initial conditions were given by a sum of
peakons of the form (\ref{vel-rad}) for some chosen initial
$p_i$ and $q_i$.
For a peakon collapsing to the center, which
will be described next, the boundary condition at the origin
is important.  If the PDE (\ref{radmeq}) were extended to
$r<0$, then the velocity would be an \emph{odd}
function about the origin.  In addition, when the peakon was
sufficiently close to the origin ($q_i<0.1$), the sign of its
momentum $m_r$ was reversed to begin its expansion away from the
origin.

For comparison with the simulations of the Eulerian PDE
(\ref{radmeq}), simulations of the Lagrangian ODEs
(\ref{caneqs}) were also performed.  A fourth-order Runge-Kutta
method was used to advance the system in time, and a time step
was chosen to ensure the Hamiltonian (\ref{hamcyl}) was
conserved to within 0.1\%.  The results of these simulations
agreed with those of the Eulerian PDE simulations to within
1\%.

\subsection{Bouncing off the center }
Let us first consider the case when only one peakon collapses onto
the center with the angular momentum being zero.  The Hamiltonian
in this case is,
\[ H = \half p^2 I_1(q) K_1(q)\,, \]
which can be approximated when $q \rightarrow 0$ as
\[
H = \left( \frac{1}{4}+o(q) \right)  p^2 \,.\]
Thus, the
momentum $p(t)$ is nearly constant just before the collapse time,
$t_*$, and is approximately equal to $-2\sqrt{H}$, more precisely,
\[
p =-2 \sqrt{H} + o(q).
\]
  The equation of
motion for $q(t)$ yields, $ \dot{q} =   p I_1(q) K_1(q)= -
\sqrt{ H}+o(q) $. If $q \rightarrow 0$ at $t \rightarrow t_*$, then
we necessarily have
\[
q(t)=  \sqrt{H} (t_* -t)+
+
o\big((t_* -t)^2\big)
\,,
\]
near the time of collapse $t\to t_*$.

The case of $N$ radial peakons can be considered similarly. If only one
peakon (let us say, number $a$) collapses into the center at time $t_*$, so
$q_a(t) \rightarrow 0$ as $t \rightarrow t_*$, and the motion of the
peakons away from the center is regular in some interval $(t_*-\delta, t_*+
\delta)$ (as will be the case unless a peakon-antipeakon collision occurs
during this interval), then conservation of the Hamiltonian implies
\beq
p_a^2 G(q_a,q_a) + 2 p_a A + B=2 H, \label{pmeq}
\eeq
where
\[
A=\sum_{i\neq a} p_i G(q_a,q_i),
\]
\[
B=\sum_{(i,j) \neq a} p_i p_j G(q_i,q_j)
\]
Since $q_a=min(q_1, \ldots, q_N)$, the Helmholtz Green's function in
expression (\ref{greensfcn}) implies that the quantities $A$ and $B$ are
bounded at times close to $t_*$, and that $G(q_a,q_a)$ is bounded as well.
Consequently, equation (\ref{pmeq}) implies that $p_a$ is also bounded at
times close to
$t=t_*$.

\begin{figure}
\centering
\includegraphics[scale=0.5]{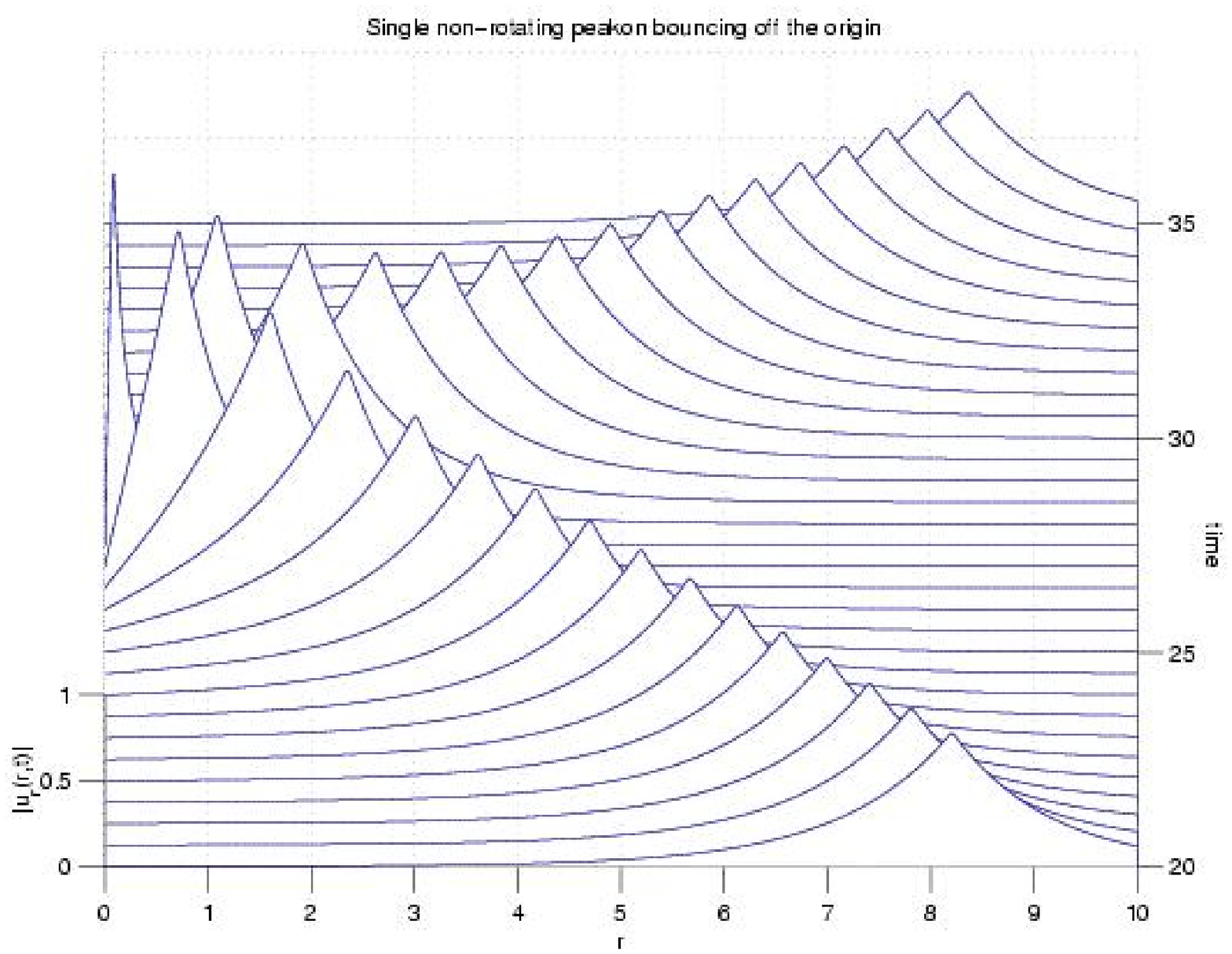}
\includegraphics[scale=0.5]{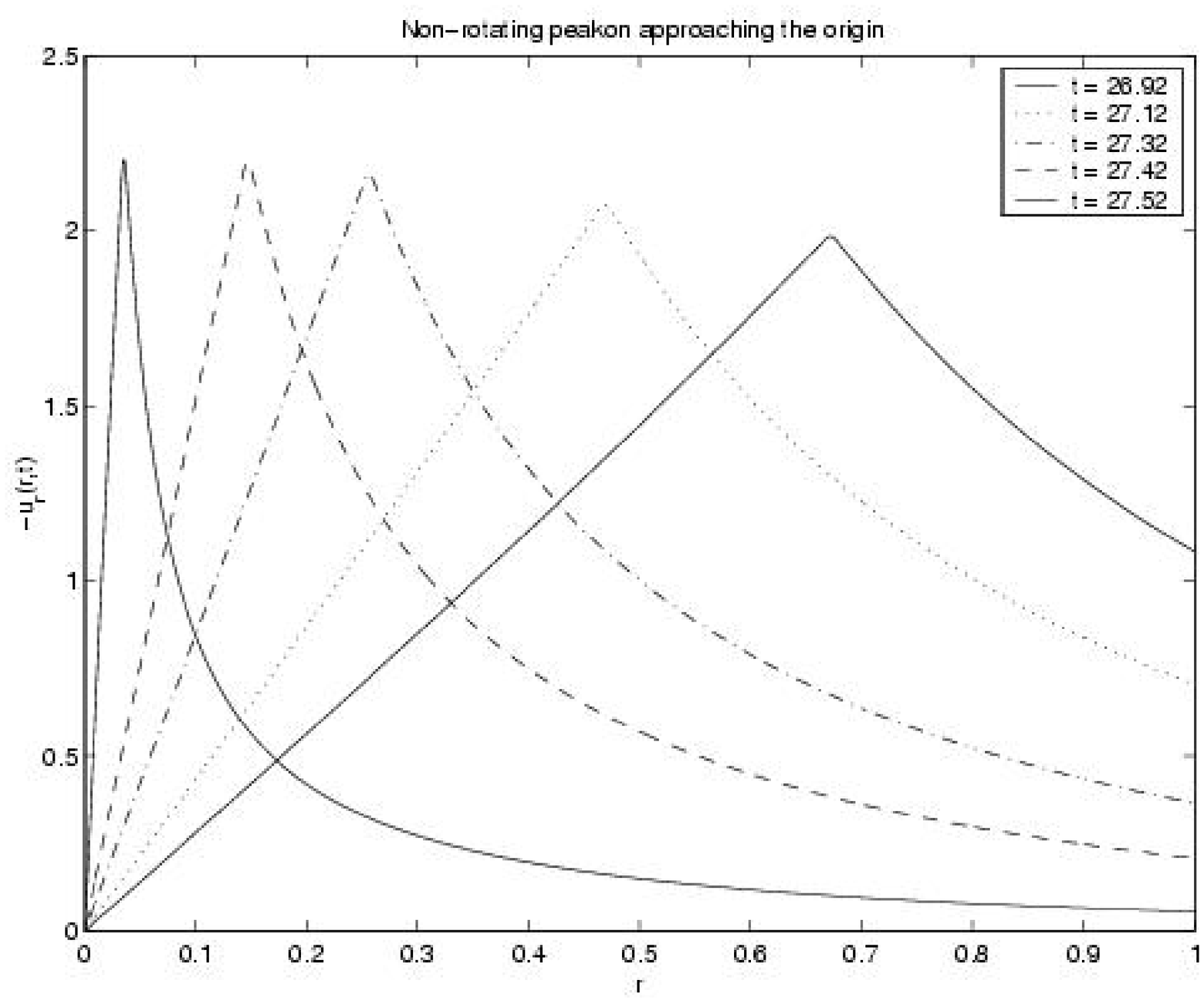}
\caption{Impact of a peakon onto the center.}
\label{impact-fig}
\end{figure}

Numerical simulations confirm our predictions: At the moment of the impact
at the center, the amplitude of the peakon remains bounded and approaches
the value of $-\sqrt{H}\approx -2.23$, as
illustrated on Fig.~\ref{impact-fig}.

The slope of the solution at the origin has to diverge. This can be seen
on the example of a single peakon
as follows.  Since
\[ \frac{\partial  u}{\partial r} \left|_{r=0} \right. =p(t) I_1'(0) K_1(q)=
\frac{1+o(q)}{2} \frac{p(t)}{q(t)}
\]
Thus, if $q(t) \rightarrow 0$ as $t \rightarrow t_*$, the slope
$\partial_r u (r=0,t)$ must diverge. Therefore, the following Proposition is true:
\begin{proposition} A radially symmetric peakon with no angular momentum, collapsing to the center,
has bounded momentum and unbounded
slope at the origin close to  the moment of collapse.
\end{proposition}

\section{Numerical Results for Rotating Peakon Circles}

\begin{figure}
\centering
\includegraphics[scale=0.4]{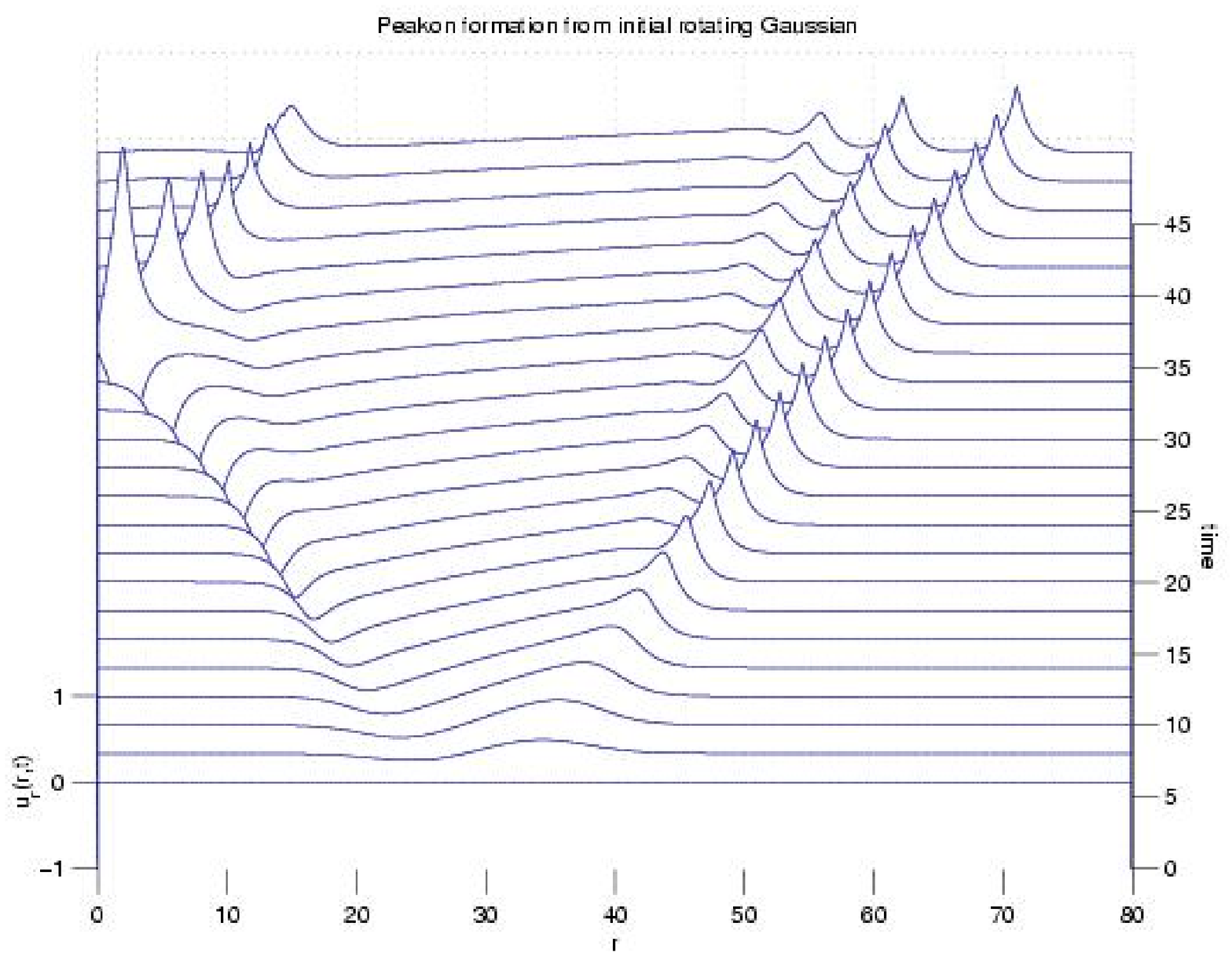}
\includegraphics[scale=0.4]{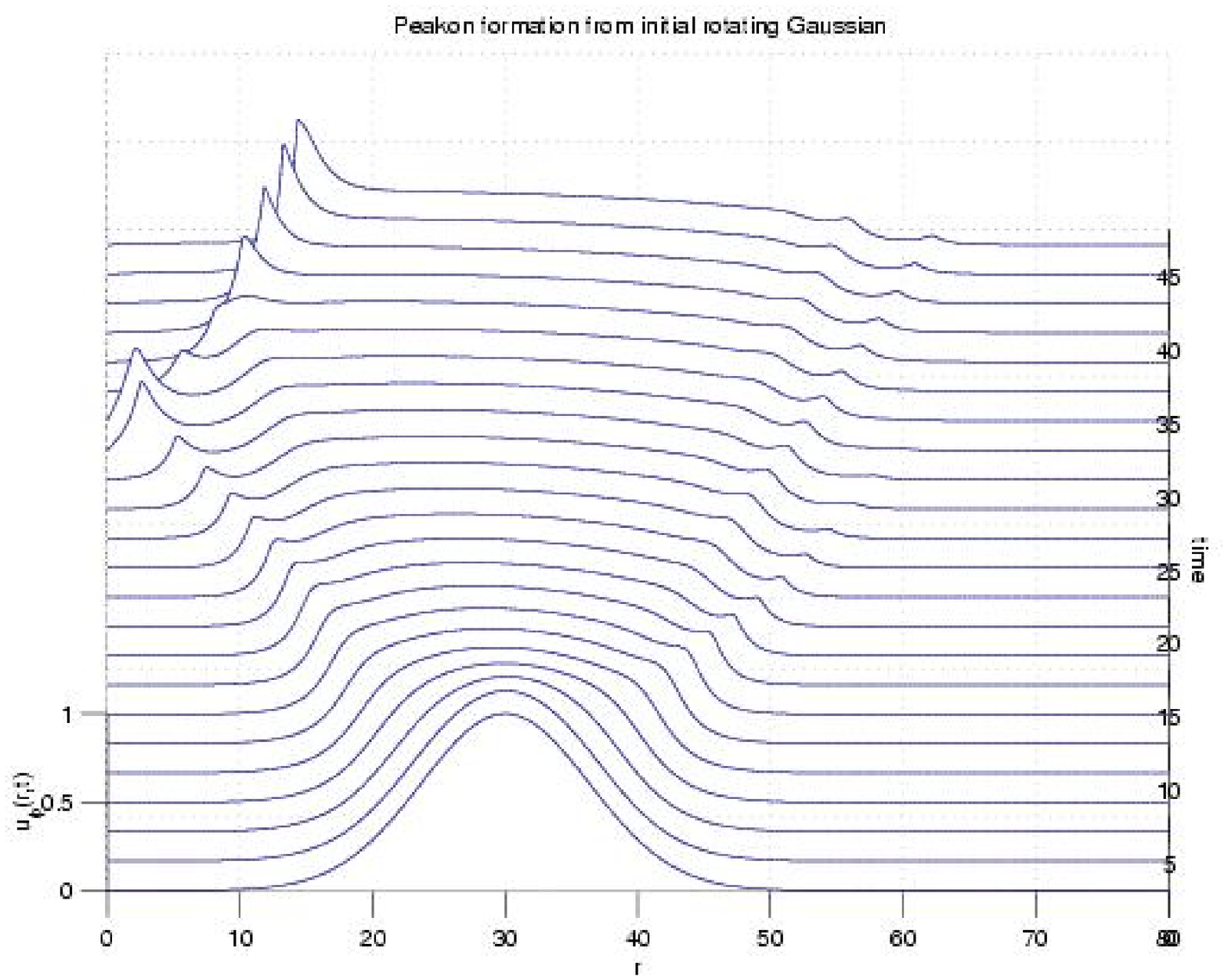}
\includegraphics[scale=0.4]{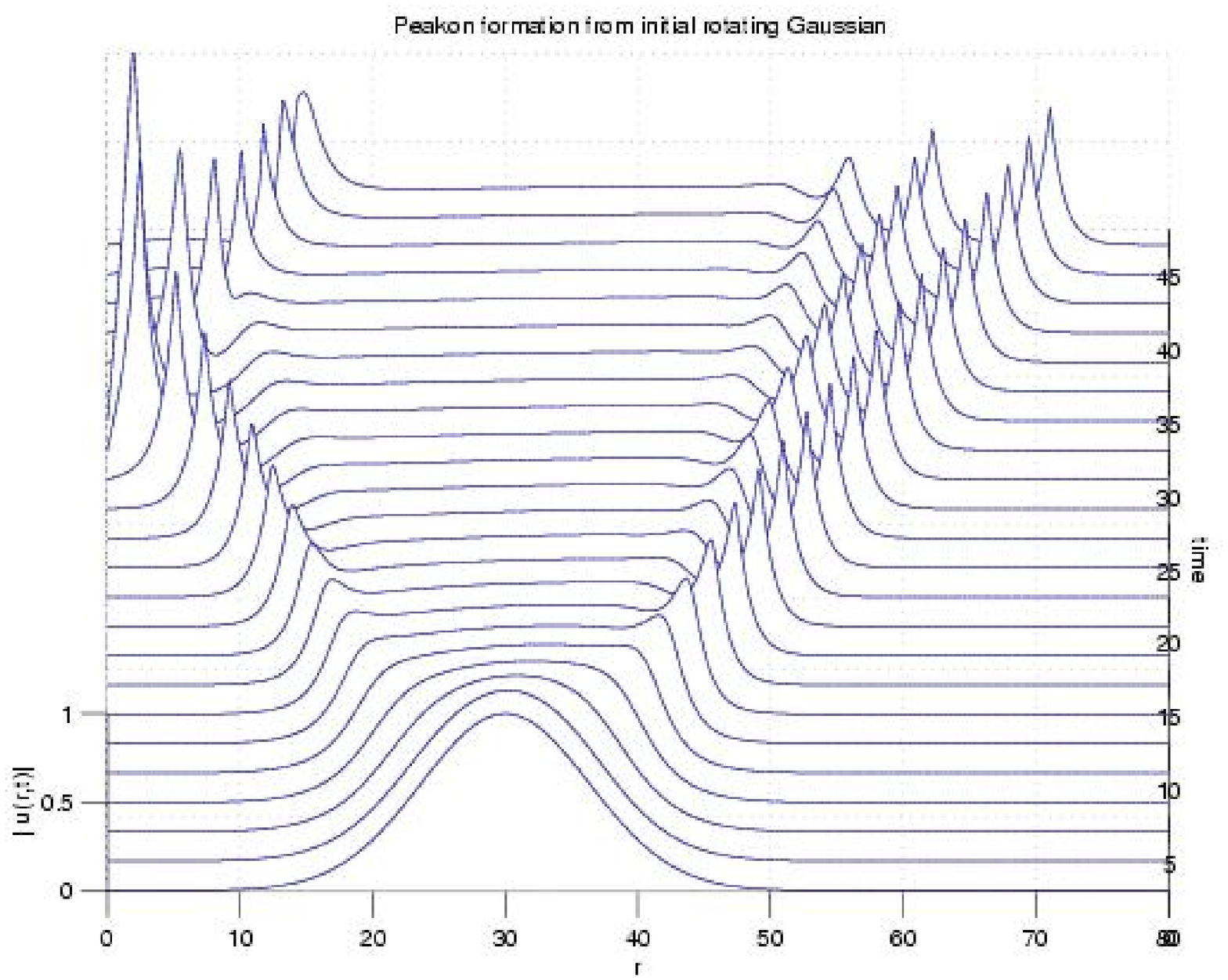}
\caption{An initial angular velocity distribution (with zero initial
radial velocity) breaks up into rotating peakons which move both
inward and outward.  The radial velocity $u_r$, the angular
velocity $u_\phi$, and the velocity magnitude $|\mathbf{u}|$ are
all shown.}
\label{rot-ivp}
\end{figure}

Simulations of rotating peakons were performed for the Eulerian system
of PDEs (\ref{mr-intro},\ref{mphi-intro}) using the same numerical
methods as those used for non-rotating peakons.  Fig.~\ref{rot-ivp}
shows the results of an initial value problem simulation when $u_r$ is
initially 0 and $u_\phi$ is initially a Gaussian
function.  Radial velocity in both directions is almost immediately
generated,
and rotating peakons soon emerge moving both inward and outward but
all rotating in the same direction.  A rotating peakon approaches the
center but turns around before reaching the origin.

\begin{figure}
\centering
\includegraphics[scale=0.5]{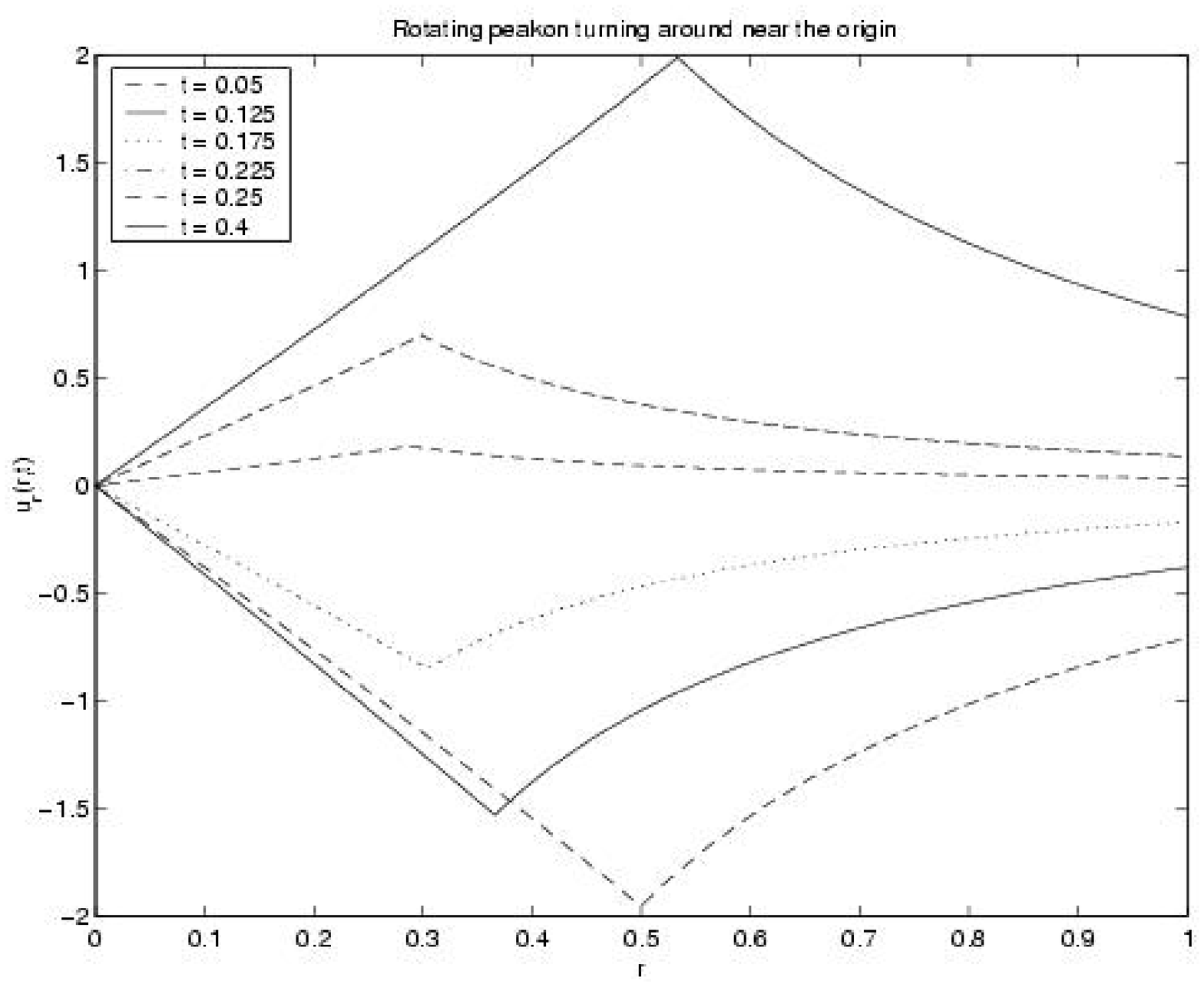}
\includegraphics[scale=0.5]{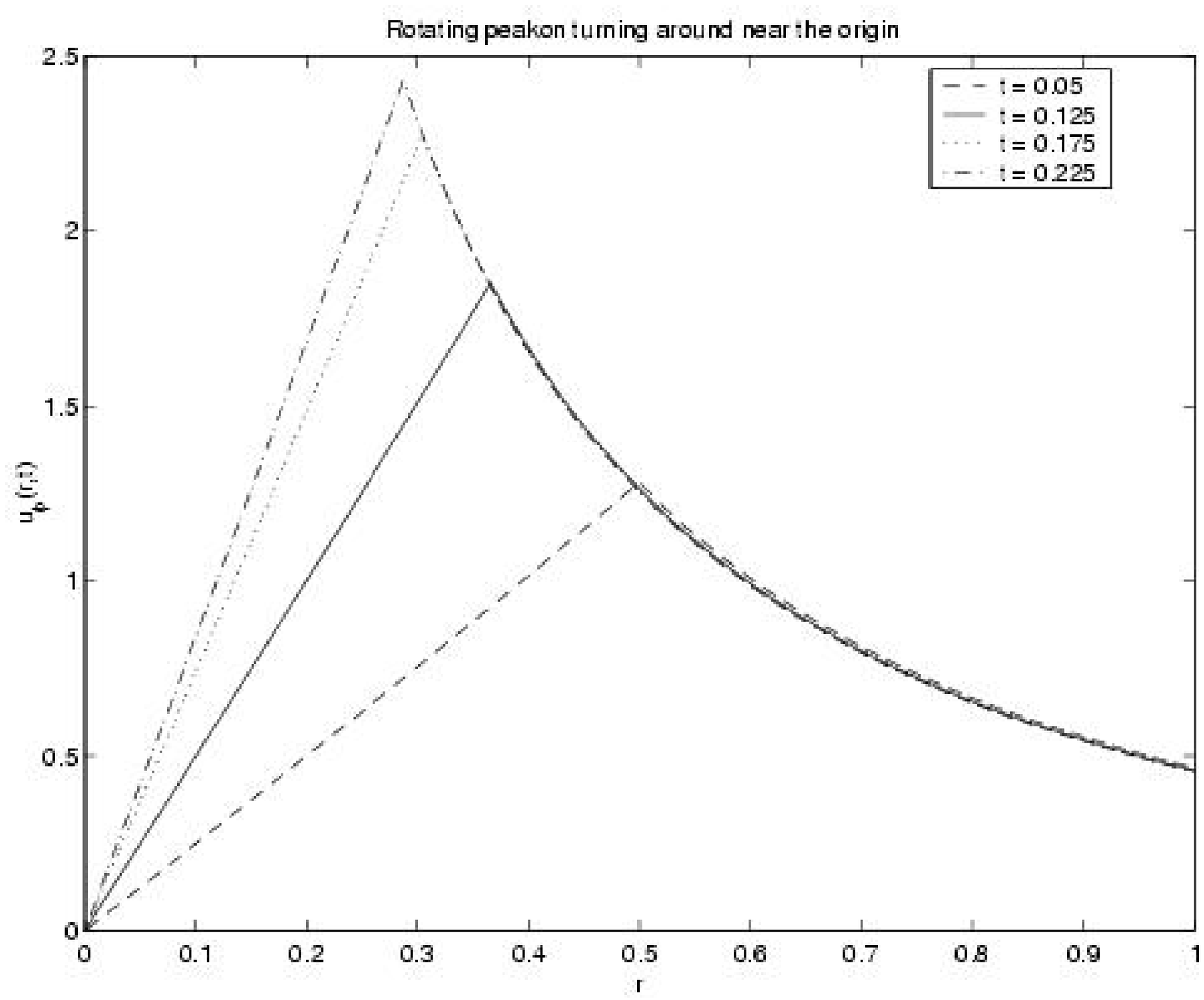}
\caption{A rotating peakon turns around near the origin.  The
radial velocity changes sign as the angular velocity reaches
a maximum.}
\label{rot-turn}
\end{figure}

Fig.~\ref{rot-turn} shows a rotating peakon as it approaches the
origin. A sort of angular momentum barrier is reached, and the
peakon turns around and moves away from the origin.  Thus, a
peakon's behavior as it approaches the origin is reminiscent of
Sundman's Theorem; if a peakon has nonzero angular momentum, then
a full collapse to the origin will not occur. This result can be
understood as follows. For a single rotating peakon, the
Hamiltonian (\ref{hamcyl}) becomes
\[
H= \left( p^2 + \frac{M^2}{q^2} \right) G(q,q)
\]
From the theory of Bessel functions we know that $G(q,q) \rightarrow 1/2$ when $q \rightarrow 0$,
and $G(q,q)>0$ for all $q$. Moreover, it can be shown that $G(q,q)$ is strictly decreasing with increasing
$q>0$, so we conclude that  $G(q,q)> G(q_0,q_0)$ if $0<q<q_0$. Thus, when $0<q<q_0$,
\[
H \geq \frac{M^2}{q^2} G(q_0,q_0)
\]
so
\beq
\label{minq}
q \geq \mbox{max} \left( q_0, M \sqrt{ \frac{ G(q_0,q_0)}{H}} \right)
\eeq
The estimate (\ref{minq}) provides the lower bound $q(t)$ can reach in the process of evolution
for each value of the parameter $q_0$. This estimate can be further optimized as follows.
Since $G(q_0,q_0)$ is strictly decreasing starting with the value of $G(0,0)=1/2$, there is $q_*(M,H)$ such that
\beq
q_*=M \sqrt{\frac{G(q_*,q_*)}{H}}
\label{qopt}
\eeq
for each value of $M,H>0$. Then, $q \geq q_*(M,H)$ is the desired optimal
estimate. We can summarize this result in
\begin{proposition}
Consider a rotating radial peakon with given values $M,H>0$.
If we define $q_*(M,H)$ by (\ref{qopt}), then in the process of evolution this peakon cannot be closer
than $q_*(M,H)$ to the origin.
\end{proposition}
{\em Note} One can show that $q_*(M,H) \sim M/ \sqrt{2 H}+O\left( M^3 \right)$ for $M \rightarrow 0$.

\section{Conclusions}

The momentum map (\ref{m-ansatz-intro}) for the action of diffeomorphisms
on (closed) curves in the plane was used to generate invariant manifolds of
measure-valued solutions of the Euler-Poincar\'e equation
(\ref{VectorEPeqn}) for geodesic motion on the diffeomorphism group. The
$H^1$ norm of the fluid velocity was chosen for the kinetic energy of a
class of solutions whose momentum support set was a family of
$N$ closed curves arranged as concentric circles on the plane. These
finite-dimensional invariant manifolds of cylindrically symmetric solutions
generalized the $N-$peakon (soliton) solutions of the Camassa-Holm equation
from motion of points along the real line, to motion with cylindrical
symmetry. This cylindrical motion included rotation, or, equivalently,
circulation, which drives the radial motion. The momentum map with non-zero
circulation for these concentric circles yielded a generalization of the
circular CH peakons that included their rotational degrees of freedom.
The canonical Hamiltonian parameters in the momentum map {\it and} solution
ansatz (\ref{spin-peakon-momentmap}) for the concentric rotating circular
peakons provided a finite-dimensional Lagrangian description in cylindrical
symmetry of the flow governed by the Eulerian EP partial differential
equation for geodesic motion (\ref{VectorEPeqn}). Numerically, we studied
the basic interactions of these circular peakons amongst themselves, by
collisions and by collapse to the center, with and without rotation.

The main conclusions from our numerical study were:
\begin{itemize}
\item
Collapse to the center without rotation occurs with bounded canonical
momentum and with vertical radial slope in velocity at r = 0, at the instant
of collapse.
\item
 For nonzero rotation, collapse to the center cannot occur and  the slope at $r=0$ never becomes
 infinite.
\end{itemize}
The main questions that remain are:
\begin{itemize}
\item
Numerical simulations show that near vertical or vertical slope occurs at
head-on collision between two peakons of nearly equal height. A rigorous proof of this fact is still missing.
\item
Is the motion integrable on our $2N$ dimensional Hamiltonian manifold of
concentric rotating circular peakons for any $N>1$ and any choice of
Green's function?
\item
How does one determine the number and speeds of the rotating circular
peakons that emerge from a given initial condition?
\item
How does the momentum map with internal degrees of freedom generalize to
$n$ dimensions?
\end{itemize}
All of these challenging problems are beyond the scope of the present paper
and will be the subjects of future work.
\rem{
The last problem seems to be the most challenging one. It is our belief
that the problem of $n$ dimensions can be understood by considering the motion of
a rotating spherical peakon in three dimensions. The main difficulty lies in the
fact that the turns along the spherical angles do not commute, which is the reason why
a spherical generalization of the ansatz (\ref{mvcyl}) will not work. Consider, for example,
a momentum filament of initially spherical shape which at time $t=0$ is rotating about the $z$ axis.
It is clear that the subsequent evolution will keep the solution invariant with respect to rotation
about the $z$ axis, but the spherical shape of the filament will be broken. Our conjecture
is that the sphere will split up into a set of
generalized rotating spherical peakons which will move along  $z$-axis in addition to the
$(r, \phi)$ dynamics considered here. The exact description of this
solution will be the subject of future work.
}

\section*{Acknowledgements}
We are enormously grateful to our friends and colleagues for their
scientific suggestions and help during the course of this work. We are
especially grateful to Martin Staley for his help and cooperation in
producing first class numerics for Figure \ref{iw-martin}. We are also
grateful for support and hospitality. D.D.H. is grateful for support by US
DOE,  from contract W-7405-ENG-36 and from US DOE, Office of Science
ASCAR/MICS. V. P. acknowledges the hospitality of the Los Alamos National
Laboratory and Center for Nonlinear Studies, where he was able to spend one
day a week during academic year 2002-2003. V. P. was partially supported by
Petroleum Foundation Research Grant Number 40218-AC9. S. S. acknowledges
summer
student support at the Theoretical Division at Los Alamos National
Laboratory.

\section{Appendix: Extension of one-dimensional peakons}
\rem{VP:  new appendix. For some reasons, it does not show on Table
of Contents. Can you fix it? }
In this appendix, we show how to obtain the momentum line peakons, which
are generalizaion of
the line peakons.
The standard ansatz  for the regular one-dimensional peakon is
\beq
m(x,t)=\sum_{i=1}^N p_i(t) \delta(x- q_i(t))
\,,
\label{m1D}
\eeq
where $m$ satisfies the one-dimensional version of (\ref{VectorEPeqn}). We
propose the following extension of these solutions:
\beq
\mathbf{m}(x,t)=\sum_{i=1}^N \left( p_i(t) \mathbf{\hat{x}} + v_i(t)
\mathbf{\hat{y}} \right) \delta(x- q_i(t))
\,,
\label{mv}
\eeq
where $\mathbf{\hat{x}},\,\mathbf{\hat{y}}$ are unit vectors in $x,y$
directions, respectively. The solution lives on line filaments, which are
parallel to the $y$ axis  and propagate by translation along the $x$ axis.
However, the $y$ component of momentum now has a non-trivial value. Such
solutions represent momentum lines which propagate perpendicular to the
shock's front and ``slide" parallel  to the front, moving surrounding
`fluid' with it. Upon substituting (\ref{mv})  into equations of motion
(\ref{VectorEPeqn}), we see that the $x$ and the $y$ components of
(\ref{VectorEPeqn}) both give the same equation of motion for $q_i(t)$:
\[
\dot{q}_i(t)=\sum_j p_j G(q_i,q_j)
\,.
\]
This compatibilty is what makes the factorized solution (\ref{mv}) possible.
The equation of motion for $p_i$ is
\[
\dot{p}_i(t)=\sum_j (p_i p_j + v_i v_j) G\,'(q_i,q_j)
\,,
\]
and for $v_i$,
\[
\dot{v}_i=0
\,.
\]
Thus, $v_i$ can be considered as a set of parameters. The $(p_i,q_i)$ still
satisfy Hamilton's canonical equations, with Hamiltonian now given by,
\beq
H=\frac{1}{2} \sumijn (p_i p_j+ v_i v_j)\,  G(q_i,q_j)
\,.
\label{Hamv}
\eeq
Finally, the ``angle'' variables $y_i(t)$ conjugate to $v_i(t)$ with
canonical Poisson bracket $\{y_i\,,\,v_j\}=\delta_{ij}$ satisfy:
\[
\dot{y}_i(t)=\sum_j v_j \, G(q_i,q_j)
\,.
\]

\end{document}